\documentclass[]{aa}
\usepackage{graphicx}
\usepackage{latexsym}
\usepackage{amssymb}
\usepackage{natbib}
\bibpunct{(}{)}{;}{a}{}{,}
\usepackage{txfonts}
\def\teff{\ifmmode T_{\rm eff} \else $T_{\mathrm{eff}}$\fi}

\def\ltsima{$\buildrel<\over\sim$}
\def\lsim{\lower.5ex\hbox{\ltsima}}

\newcommand{\hii}{H~{\sc ii}}
\newcommand{\ha}{\ifmmode {\rm H}\alpha \else H$\alpha$\fi}
\newcommand{\hb}{\ifmmode {\rm H}\beta \else H$\beta$\fi}
\newcommand{\lya}{\ifmmode {\rm Ly}\alpha \else Ly$\alpha$\fi}

\newcommand{\ebv}{\ifmmode E_{\rm B-V} \else $E_{\rm B-V}$ \fi}
\def\micron{$\mu$m}
\def\kms{km s$^{-1}$}

\def\msun{\ifmmode M_{\odot} \else M$_{\odot}$\fi}
\def\msunyr{\ifmmode M_{\odot} {\rm yr}^{-1} \else M$_{\odot}$ yr$^{-1}$\fi}
\def\zsun{\ifmmode Z_{\odot} \else Z$_{\odot}$\fi}
\def\lsun{\ifmmode L_{\odot} \else L$_{\odot}$\fi}

\def\mup{\ifmmode M_{\rm up} \else M$_{\rm up}$\fi}
\def\mlow{\ifmmode M_{\rm low} \else M$_{\rm low}$\fi}

\def\bacs{B$_{\rm 435}$}
\def\vacs{V$_{\rm 606}$}
\def\iacs{i$_{\rm 776}$}
\def\zacs{z$_{\rm 850LP}$}
\def\jnic{J$_{\rm 110W}$}
\def\hnic{H$_{\rm 160W}$}
\def\ywfc{Y$_{\rm 105}$}
\def\jwfc{J$_{\rm 125}$}
\def\hwfc{H$_{\rm 160}$}

\newcommand{\oh}{\ifmmode 12 + \log({\rm O/H}) \else$12 + \log({\rm
O/H})$\fi}
\def\hyperz{{\em Hyperz}}
\def\flyf{\ifmmode f_{\rm Lyf} \else $f_{\rm Lyf}$\fi}
\def\pz{\ifmmode P(z) \else $P(z)$\fi}
\def\ki2{\ifmmode \chi^2 \else $\chi^2$\fi}
\def\zphot{\ifmmode z_{\rm phot} \else $z_{\rm phot}$\fi}

\newcommand{\xphot}{\ifmmode x_\gamma \else $v_\gamma$\fi}
\newcommand{\xobs}{\ifmmode x_{\rm obs} \else $x_{\rm obs}$\fi}
\newcommand{\xcmf}{\ifmmode x_{\rm CMF} \else $x_{\rm CMF}$\fi}
\newcommand{\vexp}{\ifmmode V_{\rm exp} \else $V_{\rm exp}$\fi}
\newcommand{\vmax}{\ifmmode V_{\rm max} \else $V_{\rm max}$\fi}
\newcommand{\nh}{\ifmmode N_{\rm HI} \else $N_{\rm HI}$\fi}
\newcommand{\dv}{\ifmmode \Delta v({\rm em-abs}) \else $\Delta v({\rm em}-{\rm abs})$\fi}

\begin{document}
\title{On the physical properties of $z \approx$ 6--8 galaxies}

\author{Daniel Schaerer\inst{1,2}
\and
Stephane de Barros\inst{1}
}
\offprints{daniel.schaerer@unige.ch}
\titlerunning{}

\institute{
Geneva Observatory, University of Geneva,
51, Ch. des Maillettes, CH-1290 Versoix, Switzerland
\and
Laboratoire d'Astrophysique de Toulouse-Tarbes, 
Universit\'e de Toulouse, CNRS,
14 Avenue E. Belin,
F-31400 Toulouse, France
}
\date{Received 23 December 2009 / Accepted 3 February 2010}

\abstract{}{
  We analyse the spectral energy distributions (SEDs)
  of the most distant galaxies discovered with the Hubble
  Space telescope and from the COSMOS survey and determine their
  physical properties, such as stellar age and mass, dust attenuation,
  and star-formation rate.}
{
We use our SED fitting tool including the effects of nebular emission
to analyse three samples of $z \sim$6--8 galaxies
with observed magnitudes $J_{AB} \sim$ 23 to 29. Our models cover a wide
parameter space.
}
{
We find that the physical parameters of most 
galaxies cover a wide range of acceptable values.
Stellar ages, in particular, are not strongly constrained, even for objects
detected longward of the Balmer break. As already pointed out earlier,
the effects of nebular lines significantly affect the age determinations
of star-forming galaxies at $z \sim$ 6--8.
We find no need for stellar populations with extreme metallicities or
other non-standard assumptions (IMF, escape fraction) 
to explain the observed properties of faint z-dropout galaxies.
Albeit with large uncertainties, our fit results show indications
of dust attenuation in some of the $z \approx$ 6--8 galaxies, which have
best-fit values of $A_V$ up to $\sim 1$.  
Furthermore, we find a possible trend of increasing dust attenuation 
with galaxy mass, and a relatively large scatter in specific star-formation 
rates, SFR/$M_\star$.
}
{The physical parameters of very high-$z$ galaxies may be more 
uncertain than indicated by previous studies. Dust attenuation
seems also to be present in some $z \approx$ 6--8 galaxies,
and may be correlated with galaxy mass, as is also the case for
SFR.}


\maketitle
\section{Introduction}
\label{s_intro}
Finding and studying the most distant galaxies formed during the
epoch of reionisation, more recent than 1 Gyr after the Big Bang, is 
one of the challenges of contemporary observational astrophysics.
Over the past few years considerable progress has been made
in this field, pushing the observable limits beyond redshift 6
with the use of ground-based facilities and satellites.

A variety of observational programs have tried to locate $z > 6$ galaxies
using different observational techniques, mostly involving either searches for 
\lya\ emission through narrow-band filters or searches using the Lyman break 
technique -- also called the dropout technique.
These have been performed either in blank fields or in fields with galaxy clusters,
which act as strong gravitational lenses, targetting different depths and survey areas.
The objects found in this way are line emitters or Lyman break galaxies (LBGs).

Although \lya\ emitters are among the most distant galaxies
with spectroscopically confirmed redshifts \citep[see][]{Iye06,Ota08},
few have been found at $z \ga7$
\citep[see e.g.][]{Cuby07,Stark07,Willis08,Hibon09,Sobral09}.
Furthermore, because of their faintness the photometry available is 
inadequate in terms of depth to allow studies of their stellar populations.

Surveys using strong gravitational lensing were among the first
to pave the way in the of study $z>6$ galaxies
\citep[see][]{Kneib04,Pello04,Egami05,Richard06,Richard08,Bradley08,Zheng09}.
Ultra-deep fields with the Hubble Space Telescope (HST) including 
near-IR observations with NICMOS have uncovered a handful of $z \sim 7$ 
candidates in blank fields \citep{Bouwens04, Labbe06, Henry08}.
These pilot studies also showed that some of the $z \ga 7$ galaxies
could be detected at 3.6 and 4.5 \micron\ with Spitzer,
thus probing the rest-frame optical emission from these objects
\citep{Egami05,Labbe06}.

Since then, surveys of z-dropout galaxies (targeting $z \sim 7$
objects) have been extended to cover larger areas, primarily with ground-based instruments 
\citep{Mannucci07,Capak09,Castellano09,Hickey09,Ouchi09,Wilkins09},
but also with HST \citep{Henry07,Henry09,Gonzalez09}.
In most cases, however, only a few near-IR  photometric bands are 
available, providing so far information only on source counts
and luminosity functions, but precluding more detailed studies of the physical 
properties of the sources.
Notable exceptions are the work of \citet{Capak09}, who present
three bright ($J \sim 23$) $z \ga 7$ galaxy candidates from the COSMOS 2 square
degree field, and the study of \citet{Gonzalez09} finding 11
fainter (\jnic $\sim$ 26.--27.5) $z \sim 7$ galaxies in the two GOODS fields. 
Both studies benefit from a coverage including optical, near-IR, and Spitzer
bands.

Observations taken recently with the newly installed WFC3 camera 
on-board HST have just been released, resulting in publications from
four independent groups identifying faint (\jwfc $\sim$ 27--29) $z \ga 7$ 
galaxies, based on the combination of the deepest available ACS/HST and 
WFC3 data \citep{Oesch09,Bouwens09_z8,Bunker09,Mclure09,Yan09}.
While these objects are too faint to be detected at the current
limits of the deepest Spitzer images, a stack of 14 z-dropout galaxies
from \citet{Oesch09} shows tentative (5.4 and 2.6 $\sigma$) detections
at 3.6 and 4.5 \micron, respectively \citep{Labbe10}.

Given these detected $z \sim 7$ galaxies (or candidates) with
available multi-band photometry, it is of interest to
determine their physical properties such as stellar ages, reddening,
stellar masses, star-formation rates, and related properties such as
their formation redshift, specific star-formation rate, and others.
Several studies have addressed these questions using 
different modeling tools \citep[see][]{Bouwens09_betaz7,Capak09,Gonzalez09,Labbe10}.
However, some consider only special types of star-formation
histories (constant star-formation rate), or zero dust extinction, and except for 
\citet{Capak09} none of them
accounts for the effects of nebular emission (lines and continua)
present in star-forming galaxies. Neglecting the latter may
in particular lead to systematically older stellar ages,
to lower dust extinction, and differences in stellar masses, as shown
by \citet{SdB09} for $z \sim 6$ galaxies.
Furthermore, the uncertainties in the derived physical parameters
are not always determined or addressed.
Last, but not least, no ``uniform'' study of the entire data sets
of $z \sim 7$ galaxies has yet been undertaken using the 
same methodology and modeling tools.
For all these reasons, we present a critical analysis of the physical
properties of the majority of $z\sim$ 6--8 galaxies that have been 
discovered recently.

Nebular emission can significantly alter
the physical parameters of distant star-forming galaxies derived
from broad-band photometry. The main reason for this is that the
emission lines, which are invariably present in the
\hii\ regions accompanying massive star-formation, strengthen with
redshift, because their observed equivalent width scales with
$(1+z)$. Since the main emission lines are in the optical
(rest-frame) domain and few are in the UV, their presence can mimick
a Balmer break in absorption, a signature usually interpreted as an age
indicator for stellar populations \citep{Kauffmann03,Wiklind08}. 
This effect of emission lines, and to a lesser
extent also nebular continuum emission, can lead to degeneracies in
broad-band SED fits of high-$z$ galaxies as e.g., shown by
\citet{Zackrisson08} and \citet{SdB09}.
The presence of both nebular lines and continua and their contribution
to broad-band photometry is well known in nearby star-forming galaxies,
such as very metal-poor objects (e.g., I Zw 18, SBS 0335-052, and others),
blue compact dwarf galaxies and related objects 
\citep[cf.][]{Izotov97,Papaderos02,Pustilnik04,Papaderos06}.
The strongest evidence of a significant contribution of the nebular continuum 
in some nearby star-forming galaxies is the observational finding of a Balmer jump
in emission \cite[see][]{Guseva07}.
For these reasons, it is important to include nebular emission
in SED fits of distant starbursts and to examine their effect
on the derived physical properties.

In the present paper, we analyse samples of $z\sim$ 6--8 galaxies discovered 
recently. The data are compiled from the literature,
including the brightest objects from the sample of \citet{Capak09},
the ``intermediate'' sample of \citet{Gonzalez09}, and
the faintest z-dropouts recently found with the WFC3 camera.
Applying our up-to-date spectral energy distribution
(SED) fitting tool, we search in particular for possible trends
in the physical parameters of $z \sim 7$ galaxies over a range of
$\sim$ 6 magnitudes, i.e., a range of $\sim$ 250 in flux.  First results
from our analysis are presented here. A more detailed and extensive
study of the properties of z-dropout galaxies and comparisons with
objects at lower redshift will be published elsewhere.

In Sect.\ \ref{s_obs}, we summarise the galaxy sample and the SED
fitting method. In Sect.\ \ref{s_res}, we present our results for the three
subsamples. The overall results of the whole $z\sim 7$ LBG sample
and implications are discussed in Sect.\ \ref{s_discuss}, where we also 
compare our results to those for LBGs at lower redshift.
Our main conclusions are discussed and summarised in Sect.\ \ref{s_conc}. 
We assume a flat $\Lambda$CDM cosmology with $H_0=70$ \kms\
Mpc$^{-1}$, $\Omega_M=0.3$, and $\Omega_{\rm vac}=0.7$.
All magnitudes are given in the AB system.

\section{Observational data and modelling tools}
\label{s_obs}

\subsection{$z \approx 7$ galaxy samples}
To determine the physical properties of $z \approx 7$ galaxies and their
uncertainties, we chose the following three samples:

\begin{itemize}
\item Two of the three bright ($J \sim 23$) z dropout galaxies from the COSMOS survey, discovered
by \citet{Capak09}. We refer to these as the ``bright sample''.
\item The 11 \zacs\ dropout objects identified by \citet{Gonzalez09} from the HST ACS
and NICMOS data in the GOODS and HUDF fields, plus their mean SED. 
These objects typically have \jnic $\sim$ 26.--27.5, and are referred to as the 
``intermediate sample''.
\item The ``faint sample'', including 15 of the 16 \zacs\ dropout candidates
found by \citet{Oesch09} in the HUDF using the newly installed WFC3 camera of HST,
and the 15 additional objects identified as $z \sim$ 6--9 candidates by  \citet{Mclure09}.
The photometry is taken from \citet{Mclure09}.
They typically span a range from \jwfc $\sim$ 27 to 29.
We also include the stacked SED obtained by \citet{Labbe10} for 14 objects
from the \citet{Oesch09} sample, which shows tentative (5.4 and 2.6 $\sigma$) 
detections in the 3.6 and 4.5 \micron\ bands of Spitzer.
\end{itemize}

The following photometric data/filters was used for the samples:
{(1)} $i^+$, $z^+$ from SuprimeCam on SUBARU, $J$, $H$, $K$ from WIRCAM on the CFHT,
and channels 1-4 of IRAC/Spitzer for the \citet{Capak09} sample.
Since object 2 is detected in the $i+$ band and at 24 \micron, and its SED 
indicates a low redshift \citep[$z\sim 1.6$,][]{Capak09}, we exclude it from
our analysis.
{(2)} \bacs, \vacs, \iacs, \zacs\ filters of ACS/HST, \jnic\ and \hnic\
of NICMOS/HST, $Ks$, and channels 1-2 of IRAC/Spitzer for the \citet{Gonzalez09} objects.
We adopted the properties of the $Ks$ filter of ISAAC/VLT for all objects.
{(3)} \bacs, \vacs, \iacs, \zacs\ filters of ACS/HST, \ywfc, \jwfc, \hwfc\
filters of WFC3/HST, and channels 1-2 from IRAC/Spitzer for the faint sample.
The original photometry from the respective papers was adopted.

Except for the 3 objects in the bright sample for which one spectral line
was found for each of them, no spectroscopic redshifts are available 
for these objects. We therefore treat the redshift as a free parameter
for all objects.

\subsection{SED fitting tool}
To analyse the broad-band photometry, we use a modified version of the \hyperz\
photometric redshift code of \citet{hyperz} described in
\citet{SdB09}. 
The main improvement with respect to both earlier versions and other
SED fitting codes is the treatment of nebular emission (lines 
and continua), which can have a significant impact on the broad-band 
photometry of high redshift galaxies and hence their
derived properties \citep[see][]{SdB09}.
We use a large set of spectral templates
\citep[primarily the GALAXEV synthesis models of ][]{BC03}, 
covering different metallicities and a
wide range of star formation (SF) histories (bursts, exponentially
decreasing, or constant SF), and we add the effects of nebular
emission. Models with a more sophisticated description of 
stellar populations, chemical evolution, dust evolution,
and different geometries \citep[see e.g.,][]{Schurer09}
are not used, given the small number of observational constraints.

We adopt a Salpeter IMF from 0.1 to 100 \msun,
and we accurately consider the returned ISM mass from stars.
Nebular emission from continuum processes and lines is added to the predicted spectra 
from the GALAXEV models, as described in \citet{SdB09}, proportional
to the Lyman continuum photon production.
The relative line intensities of He and metals are taken from \cite{Anders03}, 
including galaxies  grouped into three metallicity intervals covering 
$\sim$ 1/50 \zsun\ to \zsun. 
Hydrogen lines from the Lyman to the Brackett series are included 
with relative intensities given by case B.
Our treatment therefore covers the main emission lines of H, He, C, N, O, and S from the 
UV (\lya) to 2 \micron\ (rest-frame), necessary for fitting the SED of
galaxies at $z>4$ up to 10 \micron\ (IRAC channel 4).

The free parameters of our SED fits are:
the redshift $z$,
the metallicity $Z$ (of stars and gas),
the SF history described by the timescale
$\tau$  (where the SF rate is SFR $\propto \exp^{-t/\tau}$),
the age $t$ defined since the onset of star-formation, 
the extinction $A_V$ described here by the Calzetti law \citep{Calzetti00},
and whether or not nebular emission is included.
In some cases, we exclude the \lya\ line from the synthetic 
spectra, since this line may be attenuated by radiation transfer processes
inside the galaxy or by the intervening intergalactic medium.

Here we consider
$z \in [0,12]$ in steps of 0.1, three metallicities $Z/\zsun=$1, 1/5, 1/20, 
$\tau=$ 5, 7, 10, 30, 50, 70, 100, 300, 500, 700, and 1000 Myr in addition to 
bursts and SFR=constant, 
ages up to the Hubble time, and $A_V=$ 0--2 (or 4) mag in steps of 0.2.
In general, the combination of all parameters leads to $\sim 3 \times 10^6$ 
models for each object. 

Non-detections are included in the SED fit with \hyperz\ by setting the
flux in the corresponding filter to zero, and the error to the $1 \sigma$
upper limit. 
For all the above combinations we compute the $\chi^2$ and the scaling 
factor of the template, which provides information about the SFR and $M_\star$,
from the fit to the observed SED.
Minimisation of \ki2\ over the entire parameter space yields the best-fit 
parameters.

To illustrate the uncertainties in the resulting fit parameters,
we examine the distribution of \ki2\ across the entire parameter
space. To determine confidence intervals from the 
\ki2\ distribution, the degree of freedom must be known to
determine the $\Delta \ki2$ values corresponding to different 
confidence levels, or Monte Carlo simulations must be carried out.
In any case, the photometric uncertainties, typically taken 
from SExtractor, would also need to be examined critically, since 
these may be underestimated, and since errors in the relative
photometric calibration between different telescopes/instruments,
which affect SED fits, are usually not taken into account.
We chose to plot the 1D \ki2\ distribution for the parameter
of interest, marginalised over all other parameters, so that the reader
is able to appreciate these distributions.
Illustrative confidence intervals are determined by assuming
$\Delta \ki2 \approx 1$, the value for one degree of freedom.
This should provide a lower limit to the true uncertainties.
More quantitative estimates of the uncertainties will be given in
a subsequent publication, which will include the analysis of a larger sample 
of LBGs at different redshift.

\section{Results}
\label{s_res}

\subsection{Photometric redshifts: overview of the full sample}
\label{s_zphot}
\begin{figure}[tb]
\includegraphics[width=8.8cm]{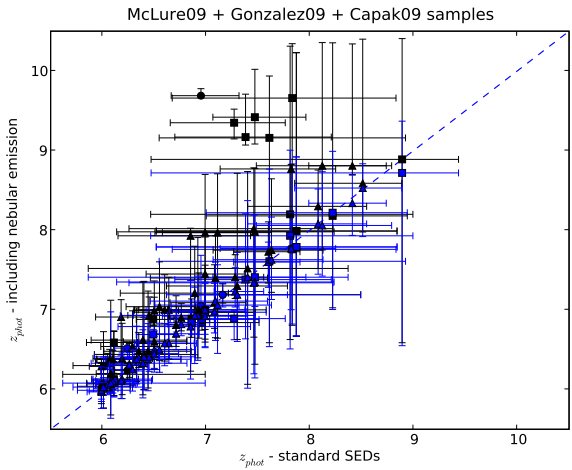}
\caption{Comparison of best-fit model photometric redshifts for the three samples using SEDs 
without (x-axis) and with nebular emission (y-axis). Circles, squares, and triangles
indicate the bright, intermediate, and faint samples, respectively. Black symbols show the
comparison with all nebular lines, i.e.\ including also \lya, blue symbols
with \lya\ suppressed. The error bars shown here denote the 68\% confidence interval
derived by \hyperz\ from the redshift probability distribution derived assuming 
$P(z) \propto \exp(-\ki2(z))$, where $\ki2(z)$ is the minimum chi-square value
over all other parameters. 
Discussion is given in text.
}
\label{fig_zphot}
\end{figure}

The photometric redshifts \zphot\ of the objects from the three subsamples
were discussed by \citet{Capak09,Gonzalez09}, and \citet{Mclure09}. Since these
authors use different spectral templates and methods, it is useful
to examine briefly the redshifts we derive from our SED fits,
and their dependence om nebular emission.
Figure \ref{fig_zphot} shows the best-fit model values for \zphot\ using
either standard templates (i.e., neglecting nebular emission), or including
nebular emission (lines and continua), and the latter but neglecting
the contribution from \lya.
Clearly, the contribution of \lya\ can lead to higher photometric
redshifts, since it can compensate for the drop of the flux
shortward of \lya, and hence lead to drop-out at higher \zphot.
With the prescription used for \lya\ in our models (i.e.\ maximum
emission according to Case B recombination) this  typically leads to $\Delta \zphot \la 1$.
In some cases, e.g., for 4 objects from the \citet{Gonzalez09} sample,
the shift is larger. The reason for this large shift is the
available filter set, which include \zacs\ and \jnic\ for this sample,
whereas \ywfc, a filter that is intermediate between \zacs\ and \jwfc, is available
for the WFC3/HST (faint) subsample.
By including the nebular continuum and all spectral lines {\em except} \lya\
(blue symbols) one recovers essentially the same photometric redshifts
as with standard templates. This is expected, since the Lyman break 
--- the main feature determining \zphot\ --- can only be strongly affected 
by \lya.

We compared our photometric redshifts against objects
with known spectroscopic redshifts, where possible. For a sample of $B$, $V$, and 
$i$-dropouts from the GOODS fields, we find good agreement for the 
majority of objects using the GOODS-MUSIC photometry \citep{Santini09}. 
For this sample, spanning objects with $z_{\rm spec} \sim$ 4--6,
our results are essentially the same with/without nebular emission,
and with/without \lya. 
Since \lya\ emission may be weaker than predicted by the models,
because of the multiple scattering in the presence of dust \citep{Verh08}
and/or because of the intervening IGM, we subsequently consider 
models including all nebular lines except \lya.
It must, however, be noted that for objects with strong \lya\ emission
the true redshift may be higher than \zphot\ obtained from photometric codes 
neglecting this line.

\begin{figure}[tb]
\includegraphics[width=8.8cm]{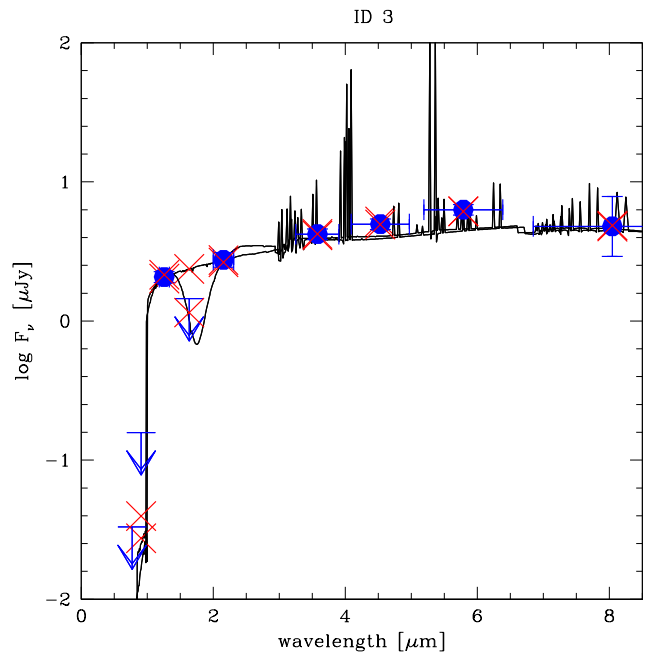}
\caption{Observed (blue points) and fitted SEDs (solid lines) of object 3
from \citet{Capak09} using our spectral templates with nebular 
emission. \lya\ emission has been supressed here.
Two best-fit models are shown comparing the standard (Calzetti) attenuation law 
for $A_V=1.2$ with the Galactic extinction law of \citet{Seat79}. Using the latter 
with $A_V=1.6$ reproduces nicely the observed dip in the H-band.
The errorbars of the observed wavelength indicate the surface of the normalised
filter transmission curve. Upper limits in flux indicate $1 \sigma$ limits.
Red crosses show the synthesised flux in the filters.}
\label{fig_capak_3}
\end{figure}

\subsection{Bright sample}
\label{s_bright}
Our SED fits for these objects yield results (redshift probability distributions
and physical parameters) broadly in agreement with \citet{Capak09}, which is 
unsurprising since these authors also include nebular lines in their
analysis using the Le Phare code.
Since we include SEDs spanning metallicities from \zsun\ to 1/50 \zsun\
(in contrast to \zsun\ only), we obtain a wider range of acceptable 
fit parameters.

For example, we find evidence of significant reddening in object 1,
with $A_V \sim$ 1--2.6, in agreement with the best-fit value of $A_V=1.2$ 
given by \citet{Capak09}. The corresponding \ki2\ distribution 
is shown in Fig.\ \ref{fig_capak_av}.
The situation is similar for object 3, although for 
more moderate extinction ($A_V \sim$ 0.6--1.6).

The corresponding range of ages, SFRs, and stellar masses for both
objects, obtained with and without nebular emission, are illustrated in 
Figs.\ \ref{fig_capak_age} to \ref{fig_capak_sfr}.
Approximately (within $\Delta \ki2 \approx$ 1-2.3) 
object number 1 (3) has thee best-fit model parameters 
$t \sim$ 0--30 (10-200) Myr,
$\log(M_\star) \sim$ 10.8--12. (10.6--11.6) \msun,
and SFR $\sim$ 100--10$^6$ (10--10$^4$) \msunyr. 
We note that a very large range in SFR is obtained from the SED fits
since both the SF history and the extinction are kept free.
The high SFR tail is related to solutions with high extinction 
and very young populations, where the UV output per unit SFR remains
below the equilibrium value reached after typically $\ga$ 100 Myr.
For comparison, using the standard SFR(UV) calibration of \citet{Kenn98}
and assuming $z=8$, one obtains SFR$_{\rm obs} \sim$ 240 \msunyr\
without extinction correction for object 1, and SFR$_{\rm corr} \sim$ 6500 \msunyr\
adopting $A_V=1.2$ and Calzetti's attenuation law.


As noted by \citet{Capak09}, object 3 is not detected in H, possibly indicating
a dip in the flux between J and K (see Fig.\ \ref{fig_capak_3}). If real, this 
dip could be explained by the 2175 \AA\ dust absorption feature, as also mentioned
by \citet{Capak09}. Using the Galactic extinction law from \citet{Seat79}
provides excellent fits with $A_V \sim 1.6$, as shown in Fig.\ \ref{fig_capak_3}.
The possible indication for a 2175 \AA\ dust absorption feature at such a high redshift
is in contrast to evidence so far, suggesting the absence of this feature
\citep[see e.g.,][]{Maiolino04}.

\begin{figure}[tb]
\includegraphics[width=8.8cm]{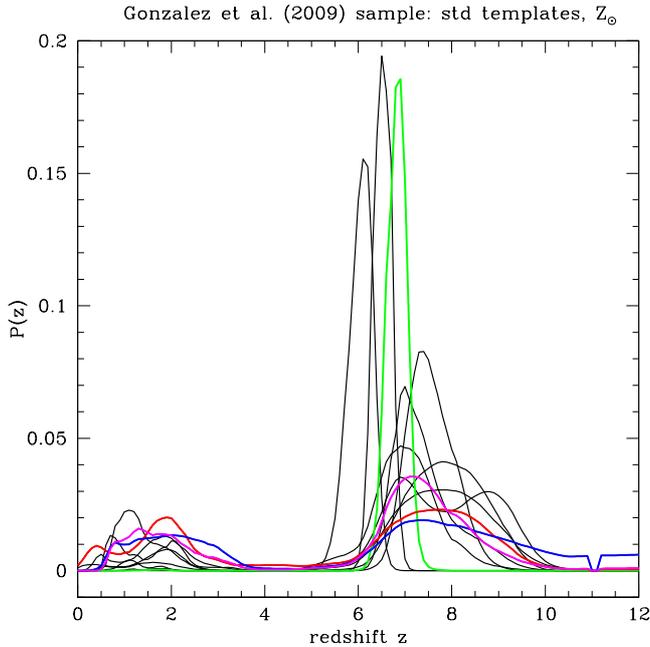}
\caption{Redshift probability distribution $P(z) \propto \exp(-\ki2(z))$
for all \citet{Gonzalez09} objects as derived from \hyperz\ including our standard 
templates. Coloured lines indicate the mean SED (green), and selected objects
discussed in the text (GNS-zD2, GNS-zD3, GNS-zD4 in red, blue, magenta). 
}
\label{fig_pz_gonz}
\end{figure}

\begin{figure}[tb]
\includegraphics[width=8.8cm]{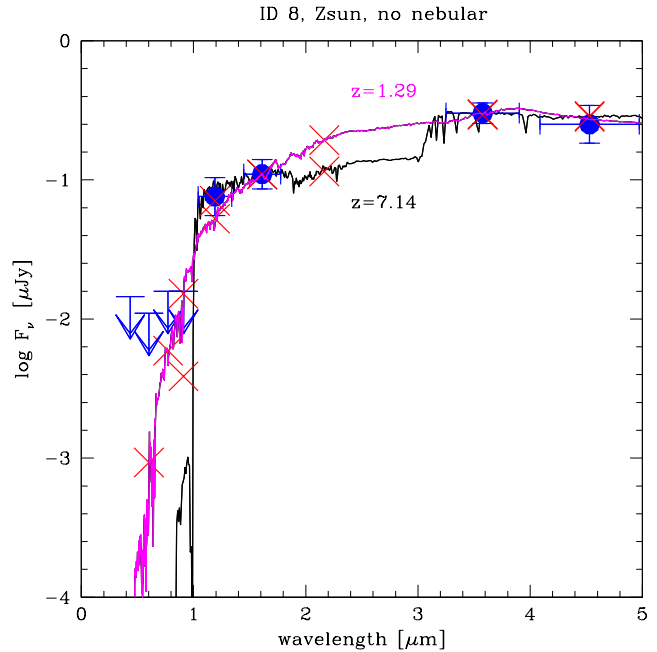}
\caption{
Observed (blue points) and fitted SEDs (solid lines) of object 8
(GNS-zD4) from \citet{Gonzalez09} using standard Bruzual \& Charlot solar metallicity 
models. The best-fit ($\chi^2=0.3$) is at $z=7.14$ 
with a stellar population of $\sim 130$ Myr (black line), the secondary solution 
($chi^2=2.1$) is a 4.5 Gyr old population at $z=1.29$ (magenta).
The errorbars of the observed wavelength indicate the surface of the normalised
filter transmission curve. Upper limits in flux indicate $1 \sigma$ limits.
Red crosses show the synthesised flux in the filters.}
\label{fig_sed_gonz8}
\end{figure}

\begin{figure}[tb]
\includegraphics[width=8.8cm]{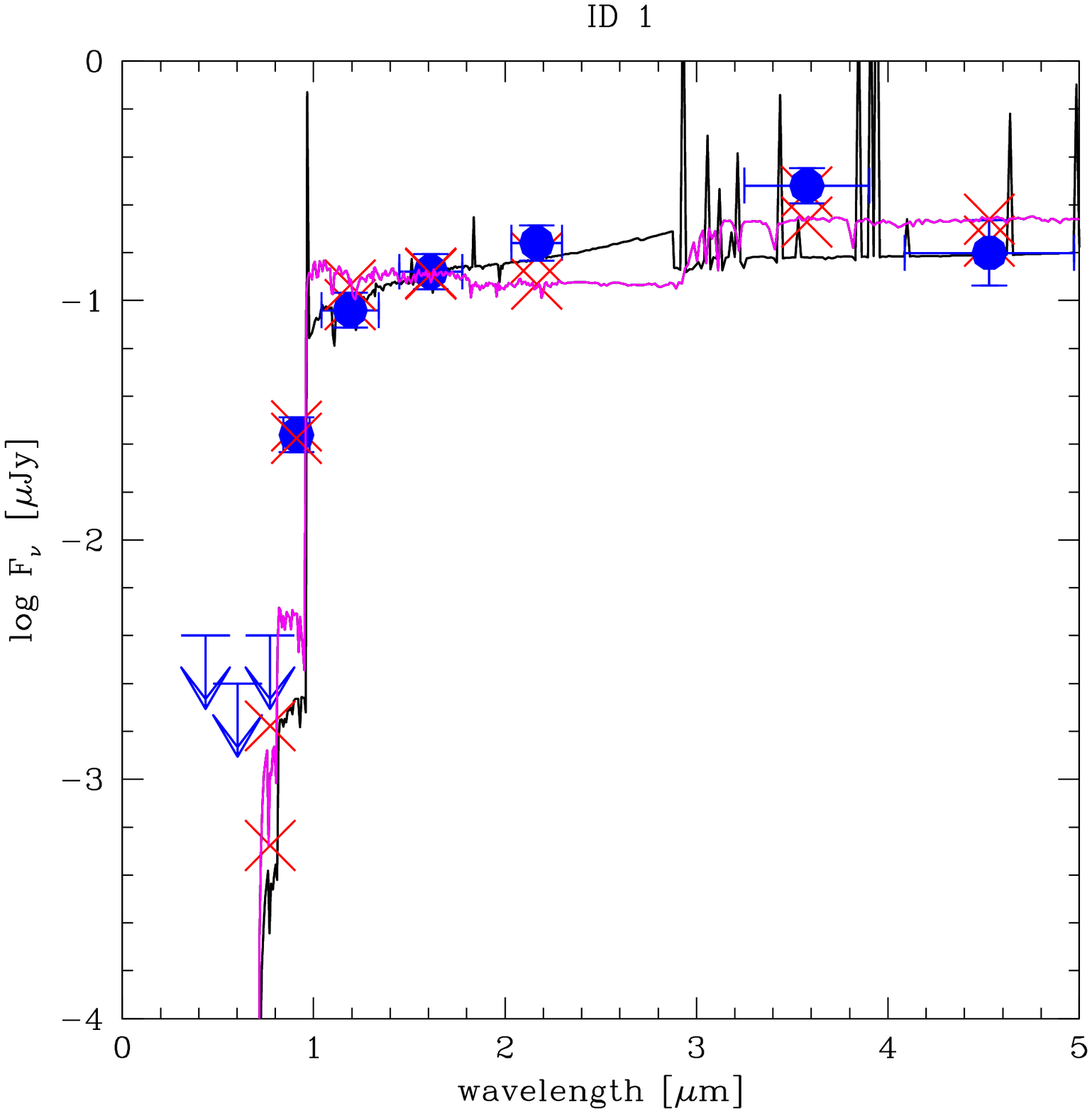}
\caption{
Observed (blue points) and best-fit SEDs (solid lines) of object 1
(UDF-640-1417) from \citet{Gonzalez09} using standard Bruzual \& Charlot solar metallicity 
models.
The magenta line shows the best-fit for templates without nebular emission, and
assuming that SFR=constant and $A_v=0$ following \citet{Gonzalez09}. 
The age of the population
is found to be $\sim$ 500 Myr.
The black line show the best-fit allowing for nebular emission and arbitrary
SF histories and extinction, yielding a much younger age plus some extinction
($t=6$ Myr, $A_V=1.2$).
}
\label{fig_sed_gonz1}
\end{figure}

\subsection{Intermediate sample}
\label{s_intermediate}
The redshift probability distribution of the objects from the \citet{Gonzalez09} 
sample is shown in Fig.\ \ref{fig_pz_gonz}. All objects have a best-fit
photometric redshift at high $z$ ($z \ga$ 6--6.5). However, 6 of the 11 objects
(ID 4, 5, 6, 7, 8, 10) also have an acceptable fit at low redshift ($z
\sim$ 1--2) with a probability comparable to the high-$z$ solution; 
the most unreliable objects being ID 6, 7, and 8.
The 3 brightest objects in \jnic\ (ID 1, 9, 11) favour clearly high-$z$ solutions,
since they provide the largest ``leverage'' on the Lyman-break between
\jnic\ and the optical data.

Figure \ref{fig_sed_gonz8} shows an example of an object with
both acceptable low- or high-redshift solutions of similar quality.
The observed SED is reproduced well by a low extinction,
young, starburst at high-$z$ ($z=7.14$ here) or by a 4.5 Gyr old 
stellar populations with $A_V=0.4$ at $z=1.29$.
Reducing the probability of a low-$z$ interloper would obviously 
be possible with deeper optical photometry, which would place tighter constraints
on the Lyman break. 
Deep K-band data (not available for this object) or other 
constraints on the shape of the SED between 2 and 3.6 \micron, may allow us
to distinguish between the two solutions shown here, and provide stronger 
constraints on the possible Balmer break -- hence the age -- of this 
object.

Object 1 (UDF-640-1417) from \citet{Gonzalez09} is such an example,
benefiting from deeper optical imaging and $Ks$ data, as shown in
Fig.\ \ref{fig_sed_gonz1}. The former leads to a well-defined and clearly 
most probable high-$z$ solution at $\zphot \sim$ 6.7--6.9.
However, here the observed spectral shape between the rest-frame UV and
optical range (probed by JHK and 3.6-4.5 \micron, respectively) 
may imply a degeneracy between age and extinction.
While \citet{Gonzalez09} fit this SED with a dust-free 
population of several 100 Myr age (cf.\ our magenta line), 
we obtain a tighter fit with models including nebular emission
for a young population plus dust reddening 
(typically  $t \la$ 10 Myr, $A_V \sim$ 0.8--1.2).
The distribution of \ki2\ for $A_V$ and other parameters are shown
in the Appendix (Figs.\ \ref{fig_gonz_chi2_av} to \ref{fig_gonz_chi2_sfr}).

In total, we find three objects 
(UDF-640-1417, GNS-zD5, HDFN-3654-1216)
with best-fit solutions for $A_V \approx$ 0.6--1.2 and relatively young age ($t \la 10$ Myr).
Incidentally these are the three brightest objects in \jnic, which may
suggest a trend of extinction with magnitude (cf.\ below).
However, the significance of non-zero extinction is not very high,
in particular for GNS-zD5 and HDFN-3654-1216, where the 3.6 \micron\
flux is affected by a bright neighbouring source \citep[cf.][]{Gonzalez09}.
 
Considering the entire parameter space for the whole sample 
(cf.\ Figs.\ \ref{fig_gonz_chi2_av} to \ref{fig_gonz_chi2_sfr}), we 
find that age and dust extinction of most objects are not well constrained,
and could reach from few Myr up to the age of the universe at that redshift,
and from $A_V \sim$ 0 to $\la 1.6$ mag for some objects.
In particular, the data does not allow us to conclude that these
galaxies show no sign of dust extinction.
Furthermore, their age and hence formation redshift remains poorly constrained.
The same is also true for the mean SED from \citet{Gonzalez09}, which
yields results compatible with those of the individual objects, as expected
(see thick line in the plots).
For comparison, \citet{Gonzalez09} find fairly old best-fit stellar-mass-weighted ages 
$t_w \sim 200-400$ Myr\footnote{With their definition, one has $t_w = 0.5 \times t$
for SFR=const assumed by these authors, where $t$ is our definition of the stellar age.} 
typically.
Both the assumption of SFR=const and $A_v=0$ lead to the highest age, since both
effects minimise the ratio of the rest-frame visible/UV light, the main age constraint.

Allowing for wide range of SF histories, variable extinction, and for nebular emission
yields, on average, a broader age range (between a few Myr and up to the maximum age), 
room for
extinction up to $A_V \la 1.$, stellar masses from $10^{8.5}$ to $10^{10}$ \msun, and 
SFR $\sim$ 2--100 \msunyr. In other words, the properties of the galaxies
from this sample of intermediate brightness (\jnic $\sim$ 26--28), are clearly
more uncertain than indicated by \citet{Gonzalez09}, who consider only a restricted
range of the parameter space.

\subsection{Faint sample}
\label{s_faint}

Using the first UDF observations taken with the newly installed WFC3 camera
onboard HST, four studies have identified $\sim$ 11 to 20
$z\sim7$ galaxy candidates (or \zacs\ drop-outs) \citep{Oesch09,Bunker09,Mclure09,Yan09}.
What can be said about their physical properties?

\subsubsection{Photometric redshifts}
\citet{Mclure09} had previously examined the photometric redshifts and uncertainties 
for their sample, which also covers the majority of $z \sim 7$ galaxies found
by the other groups \citep{Oesch09,Bunker09,Yan09}.
Unsurprisingly, our results using a modified version of the \hyperz\ code
also used by \citet{Mclure09} and a slightly more extended template library,
confirm their findings. In particular, for the $z \sim 7$ sample of \citet{Oesch09} 
we find that their objects consistently show photometric redshifts with well-defined
probability distributions peaking between $z \sim$ 6.3 and 7.6.
For fainter z-dropouts and Y-dropouts, the photometric redshift becomes
far more uncertain, and a significant fraction of the objects could
also be low-$z$ galaxies. As already pointed out by \citet{Mclure09} and \citet{Capak09},
the depth of the optical imaging becomes the limiting factor for objects
that faint in the near-IR.

\subsubsection{UV slope}
\begin{figure}[tb]
\includegraphics[width=8.8cm]{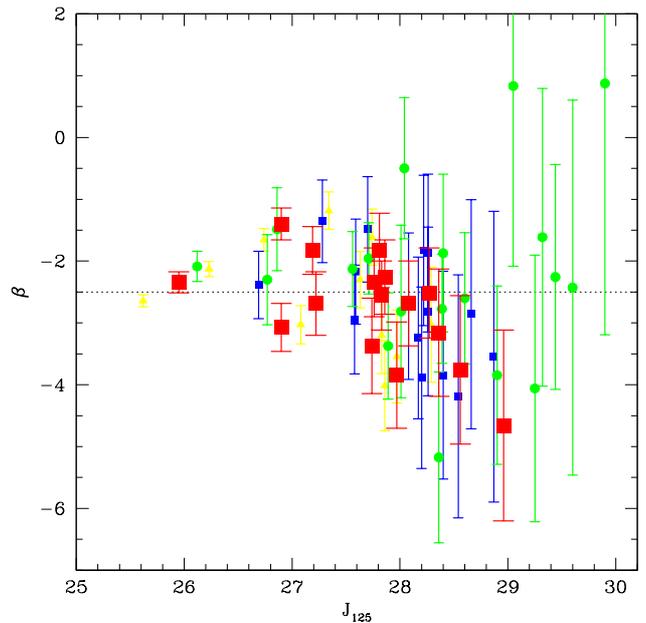}
\caption{UV slope $\beta$ between $\approx$ 1550 and 1940 \protect\AA\ computed
from $\beta = 4.29($\jwfc-\hwfc$)-2.0$ for the $z \sim 7$ galaxy candidates
from different WFC3/UDF samples. 
The photometry from various sources has been used: \citet{Oesch09} (red squares), 
\citet{Bunker09} (yellow triangles),
\citet{Yan09} (green circles), \citet{Mclure09} (blue squares, only objects
in common with Oesch et al.).
}
\label{fig_beta}
\end{figure}

One group pointed out that the fainter of these objects
had very blue UV-continuum slopes, $\beta$, indicative of 
``non-standard'' properties of these galaxies \citep{Bouwens09_betaz7}.
Their data, shown as red squares in Fig.\ \ref{fig_beta}, exhibits
a trend of decreasing $\beta$ (as estimated from their (\jwfc-\hwfc) colour) towards fainter 
magnitudes. 
From the very steep slopes (i.e., low values of $\beta \sim -3$) reached
in faint objects, \citet{Bouwens09_betaz7} claim that extremely low metallicities and 
large Lyman continuum escape fractions seem to be 
required to understand these objects, since ``standard'' evolutionary synthesis models
predict minimum values of $\beta \sim -2.5$ for young stellar populations.
As the data and the errorbars from different groups plotted
in Fig.\ \ref{fig_beta} show, we cannot reach similar conclusions, 
given the uncertainties in the colour measurement used to determine $\beta$.
For the bulk of the sources, the $\beta$-slope is compatible within $1 \sigma$ with
normal values of $\beta \sim -2.5$ or flatter slopes.
Furthermore, it is unclear whether the (\jwfc-\hwfc) colour exhibits any 
systematic trend towards fainter magnitudes.
The observations do not exclude different properties such as 
extremely low metallicities and large Lyman continuum escape fractions
for some of the objects at $z \sim 7$. 
However, the low significance of these deviations do not justify 
making assumptions that differ significantly from those commonly adopted for 
the analysis of lower redshift objects.

\begin{figure}[tb]
\includegraphics[width=8.8cm]{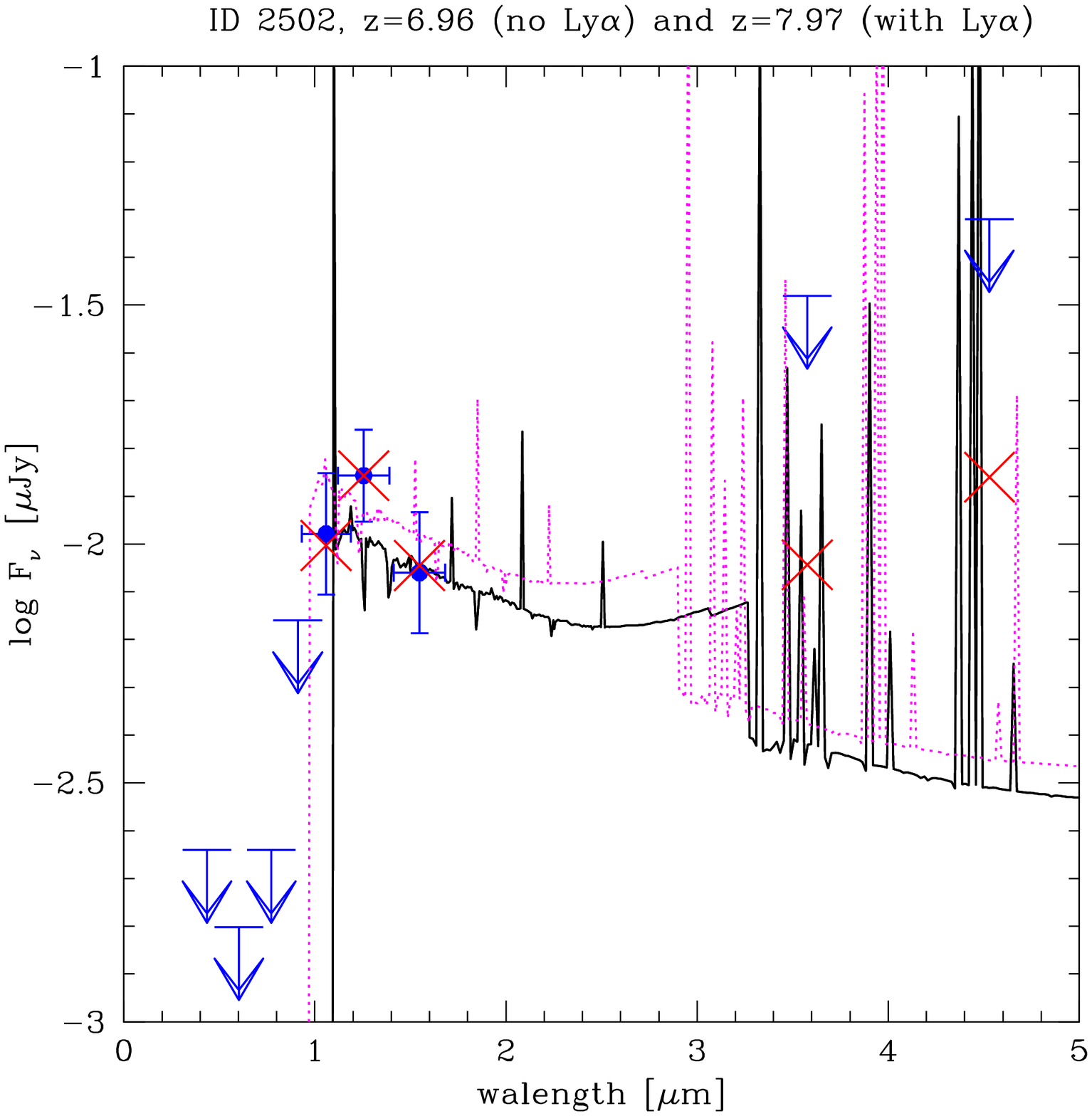}
\caption{Best-fit SEDs of object 2502 (photometry in blue symbols) of 
\citet{Mclure09} for solar metallicity models including nebular emission.
The black solid line shows the model including all lines (\lya\ in particular) 
with a best-fit \zphot=7.97. Red crosses indicate the flux in the filters
for this model.
The dashed magenta line shows the best-fit model 
excluding \lya, found at \zphot=6.96.
This object has an observed $($\jwfc-\hwfc$)=-0.51$ colour, corresponding to $\beta=-4.2$
Note the strong effect \lya\ may have on (\jwfc-\hwfc) and \zphot,
although this object is a \zacs-dropout, thought to be 
at $z \sim 7$.}
\label{fig_lyaboost}
\end{figure}

With the WFC3 filters used in this survey, it is possible
to generate unusually blue (\jwfc-\hwfc) colours in certain 
circumstances from spectral templates including nebular emission.
Such a case is illustrated in Fig.\ \ref{fig_lyaboost}, showing 
a fit to object 2502 from the \citet{Mclure09} sample.
Shown here is a model of a very young stellar population
with solar metallicity including lines and nebular continuum
emission redshifted to either $z=6.96$ (dashed, magenta line) or 
 $z=7.97$ (solid, black line). In the latter case, the strong
intrinsic \lya\ emission (with $W(\lya)^{\rm rest} \approx 200$ \AA)
boosts the flux in both \jwfc\ and \ywfc\ since
these filters overlap by $\approx 0.1$ \micron.
This provides a very blue (\jwfc-\hwfc) colour, and enough
flux in \ywfc\ to ensure that this object does not appear as a \ywfc-drop
even at $z \sim 8$. If at $z \sim 7$, as implied by SED fits excluding 
the \lya\ line (magenta line) or from simple colour-criteria 
designed to select $z \sim 7$ galaxies \citep[e.g][]{Oesch09}, 
the (\jwfc-\hwfc) colour is not affected by \lya.
However, this is not necessarily the case for all z-dropout galaxies
since strong \lya\ emission, if present, may mimic a lower 
redshift (cf.\ Sect.\ \ref{s_zphot}).
The likelyhood of this situation remains difficult to establish,
especially since \lya\ may be differentially affected by dust,
and scattered by the IGM.

\subsubsection{Age and reddening}
For the 15 objects in common between the Oesch and McLure samples, we find young
stellar populations ($t \la 10$ Myr) as best-fits, and zero extinction, except for objects
688, 835, and 1092, with $A_V \sim$ 0.2--0.6.
However, as for the objects from the intermediate sample, the distribution of \ki2\ is 
very flat (cf.\ Figs.\ \ref{fig_mclure_chi2_av} to \ref{fig_mclure_chi2_sfr}), allowing a 
wide range of extinctions ($A_V \sim$ 0--1.2), ages of $t \sim$ 0 to several 100 Myr,
stellar masses from $10^7$ to few times $10^9$ \msun, and SFRs from 0.1 (or less) to 
$\la 200$ \msunyr, for most objects in the faint sample.
The wide age range, is possible, e.g., since the upper limits at 3.6 and 4.5 \micron\
do not provide a strong enough constraint on the optical to UV flux of these faint 
objects. Given the rapid evolution with time in the mass/light ratio involved here 
(mostly the UV--optical
domain), the uncertainty in the ages translates into a large spread in
stellar masses, as show by Fig.\ \ref{fig_mclure_chi2_mass}. The wide range of 
acceptable SFR values is due to both age and SF history (parametrised
here by the e-folding timescale $\tau$) being kept free, in contrast e.g., to
commonly used SFR(UV) calibrations assuming SFR=constant and ages $t \ga$ 100 Myr.

\begin{figure}[tb]
\includegraphics[width=8.8cm]{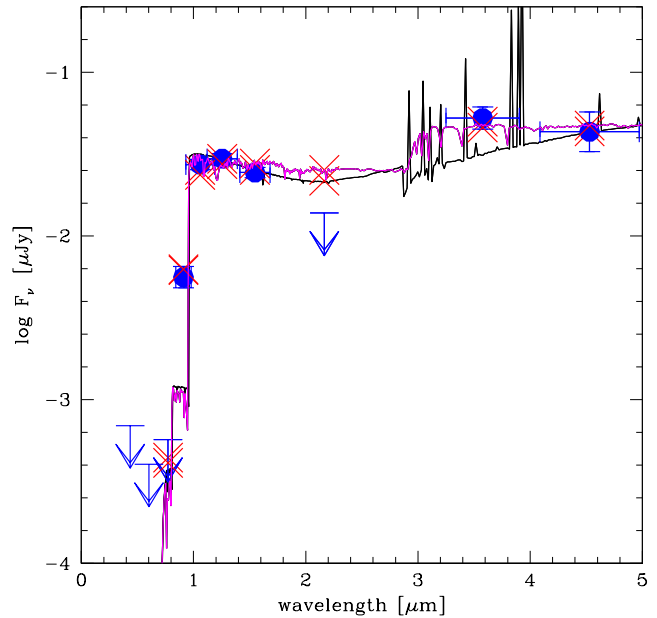}
\caption{Best-fit SEDs to the SED of the stack of 14 z-dropout galaxies
from \citet{Oesch09} measured by \citet{Labbe10}.
Photometry is shown by blue symbols; including the 1 $\sigma$ limit for Ks.
Red crosses indicate the synthetic model fluxes in the filters.
The magenta line shows the best-fit using solar metallicity templates
without nebular emission, yielding a maximum age ($t \sim 700 Myr$), $A_V=0$, 
and other parameters also similar to \citet{Labbe10}.
The black line shows the best-fit model including nebular emission (and \lya\ 
suppressed), with an age of $t=4$ Myr, $A_V=0.2$, and a stellar mass
of $M_\star \sim 5.\times 10^7$ \msun, more than 1 dex lower than the mass
corresponding to the old fit.}
\label{fig_labbe_stack}
\end{figure}

\subsubsection{Physical properties from the stacked SED}
The individual z-dropout candidates of \citet{Oesch09} and \citet{Mclure09} are 
undetected in the deep, available 3.6 and 4.5 \micron\ Spitzer images, but
\citet{Labbe10} stacked the images of 14 of the 16 z-dropout galaxies from \citet{Oesch09},
obtaining 5.4 and 2.6 $\sigma$ detections in these bands.
Fitting the SED of this stack, we find that the physical parameters
are more tightly constrained, as the thick line in Figs.\ \ref{fig_mclure_chi2_av} to 
\ref{fig_mclure_chi2_sfr} show.
Overall our best-fit values (with or without nebular
emission) are very similar to those obtained by \citet{Labbe10},
bearing the different definitions of stellar ages $t$ and $t_W$ in
mind. Furthermore, the values of the physical parameters derived for
the majority of the individual objects is compatible with the values
determined from the stack.

An uncertainty remains, however, in the age and consequently also in the stellar
mass determination. We first obtain secondary solutions with $\Delta \ki2 \sim$ 1--2
with young ages ($\sim$ 4--5 Myr) and a small extinction ($A_v \la$ 0.2--0.4)
as we consider all metallicities, both with our without nebular
emission. Furthermore, if we suppress the \lya\ line we improve the fit, leading
to a best-fit at young ages ($\sim$ 2--7 Myr).
The corresponding \ki2\ distributions illustrating these results are shown
in Figs.\ \ref{fig_stack_av} to \ref{fig_stack_sfr}, and
SED fits from these models are shown in Fig.\ \ref{fig_labbe_stack}.
This figure clearly illustrates how an apparent Balmer break can
be explained by an old population (here $t \sim$ 700 Myr) or by nebular 
emission from a younger population, as already shown by \citet{SdB09}.
A suppression of the \lya\ line is justified since \lya\
may be attenuated by radiation transfer processes inside the galaxy or by the 
intervening intergalactic medium.
In any case, neglecting nebular emission is inconsistent for spectral
templates with recent ($\la 10$ Myr) or ongoing star formation.
Finally, in comparison with the bright and intermediate samples
we may also question why the faintest $z\sim 7$ objects
should have the oldest stellar populations, whereas the ages
of brighter objects are compatible with a broad range of ages,
including young ones.

The uncertainty in the age also translates into an uncertainty in stellar mass.
Whereas the estimated average mass is $M_\star \sim (1-2)\times 10^9$ \msun\
for the old population \citep[cf.][]{Labbe10}, it is more than a factor of 
10 lower for young ages (see Fig.\ \ref{fig_stack_mass}), since nebular 
emission contributes partly to the rest-frame optical domain.
The SFR is, however, hardly affected by this uncertainty (see Fig.\ \ref{fig_stack_sfr}),
since it is more sensitive to the rest-frame UV light present in
both young and old star-forming populations.
In consequence, the specific SFR (SFR$/M_\star$) could be significantly higher
than advocated by \citet{Labbe10}.
Before performing spectroscopy for these objects -- a currently
impossible task -- to examine if the 3.6 \micron\ filter is truly
affected by emission lines as predicted by the model, the present data
does not allow us to completely rule out one or the other solution.

\section{Discussion}
\label{s_discuss}
We discuss the effects of varying the model assumptions on 
the physical parameters derived.
We then examine the main derived properties and possible correlations
among them for the ensemble of galaxies studied here, and compare
any correlations found to those of lower redshift galaxies.
Finally, we discuss some implications of our results. 

\begin{figure}[tb]
\includegraphics[width=8.8cm]{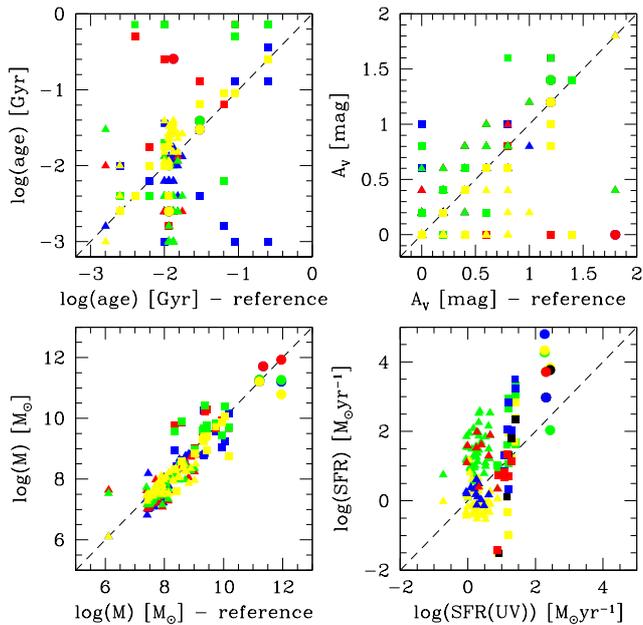}
\caption{Comparison of best-fit values for the age $t$ (upper left), extinction $A_V$ (upper right),
stellar mass $M_\star$ (lower left), and star-formation rate SFR (lower right) 
obtained with different model assumptions. The x-axis is the value obtained from
our ``reference model'' (including nebular emission but no \lya) except for the 
SFR panel; the other values
are plotted using different colours: 
models with $\tau \ge 10 Myr$ (yellow),
models with \lya\ (blue),
models without nebular emission (red), and
models without nebular emission and SFR$=const$ (green).
For the SFR comparison, we use the SFR(UV) value for the x-axis, and black
symbols for the reference model.
Filled circles, square, triangles show the bright, intermediate, and faint
samples. The dashed line is the one-to-one relation.
See discussion in text.}
\label{fig_compare}
\end{figure}

\begin{figure}[tb]
\includegraphics[width=8.8cm]{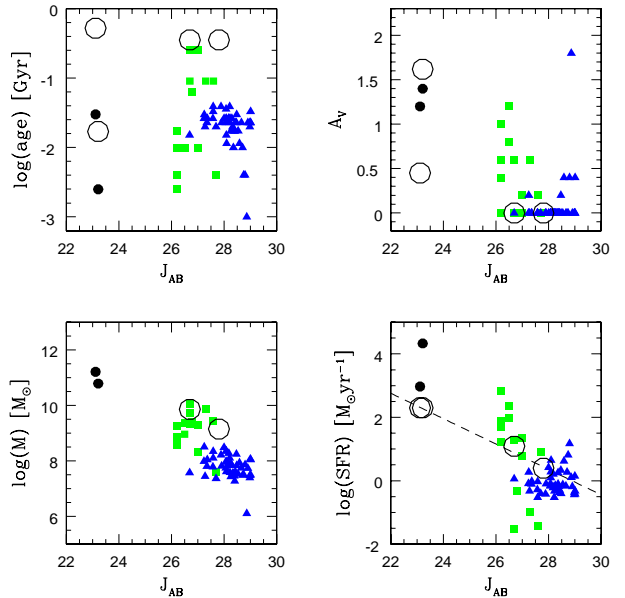}
\caption{Best-fit value  for the age (upper left), extinction (upper right),
stellar mass (lower left), and star-formation rate (lower right) as a function of 
the observed $J_{AB}$ magnitude. All best-fit values are taken from the 
models with $\tau \ge 10 Myr$ (model 1).
Filled circles, square, triangles show the bright, intermediate, and faint
samples. Large open circles show the fit values obtained by 
\citet{Capak09}, \citet{Gonzalez09}, and \citet{Labbe10} for the 2 bright objects
and from the mean/stacked SED, respectively. 
The dashed line in the SFR--$J_{AB}$ panel shows the standard SFR(UV) line
assuming $z=7$ and no extinction for all objects. See discussion in text.}
\label{fig_overview}
\end{figure}

\subsection{Effects of varying model assumptions}
Varying the model assumptions in broad-band SED fits affects the inferred  
physical parameters of distant galaxies in ways that
have been discussed in several studies, e.g., 
by \citet{Yabe09} in quite some detail \citep[cf.\ also][ and others]{Papovich01,Sawicki07,Gonzalez09}.
The most relevant assumptions are the star-formation histories, metallicity, the inclusion
of dust extinction, and the adopted extinction law. Furthermore, the inclusion
of nebular emission and the assumptions made to do so also affect the results
as shown here and in \citep[cf.\ also][ and the latter for a discussion of the 
effect of \ha]{SdB09,Yabe09}.

The impact of different model assumptions on the best-fit parameters of our sample is shown 
in Fig.\ \ref{fig_compare}, where we plot the values from our ``reference model'' (including
nebular emission, \lya\ suppressed, all SF histories, all ages , all extinction values, and all
metallicities) on the x-axis, and the same from comparison models on the y-axis.
The models we consider here for comparison are: 
{\em 1)} models with $\tau \ge 10$ Myr (yellow),
{\em 2)} models with \lya\ (blue),
{\em 3)} models without nebular emission (red), and
{\em 4)} same as 3), but for SFR=constant only (green).
SEDs without nebular emission (3) may be taken as an extreme case of
models with a very large escape fraction ($f_{\rm esc}  \approx 1$) of ionising 
radiation from the Lyman continuum, whereas in the other cases
we implicitly assume no escape ($f_{\rm esc}  \approx 0$), maximising
thus nebular emission if young massive stars are present.
One may theoretically expect an evolution of $f_{\rm esc}$ with redshift,
galaxy mass, and other properties, although large differences remain even
between simulations \citep[see e.g.,][]{Gnedin08,Wise09,Razoumov09}.
Intermediate values of $f_{\rm esc}$ may be included in future models.

Circles, squares, and triangles in Fig.\ \ref{fig_compare} represent objects from 
the bright, intermediate, and faint samples, respectively.
In Figs.\ \ref{fig_compare} to \ref{fig_mass_sfr}, we now
include all objects of \citet{Mclure09} in the faint sample, i.e.\ 15 objects
in addition to the 15 of \citet{Oesch09}, to maximise the sample size and 
because the results with the additional sample do not show noticeable differences.
We do not show the effect of fixing metallicity, since it is small compared
to the other effects discussed here.
We now discuss the dependence of the physical parameters on these models assumptions
one-by-one.

\subsubsection{Stellar mass}
As noted and discussed by \citet{Yabe09} and others, the
stellar mass $M_\star$ is the least sensitive parameter to the model assumptions, 
especially when measurements
for the rest frame optical domain are available. This is mainly because
it is derived from the absolute scaling of the overall SED, and that the 
mass-to-light ratio in the optical does not change much with age and star-formation
history.

\subsubsection{Age}
In contrast, age is the most sensitive quantity \citep[cf.][]{Yabe09}.
From Fig.\ \ref{fig_compare} it is clear that models without nebular emission (red and green)
yield older ages in most cases, as already discussed above \citep[cf.][]{SdB09}.
By assuming constant SFR, we also tend to infer older ages, as is well known, but not in all cases (see green
symbols).
The inclusion (suppression) of \lya\ can also lead to younger (older) ages, as shown by
the blue symbols. This is the case for some objects because of the 
increase of \zphot\ (cf.\ Sect.\ \ref{s_zphot}), which
in turn demands a younger age (steeper UV spectrum) to reproduce the same observed
slope in the UV. In other words, to some extent there is also a degeneracy between age and redshift
\footnote{In addition, there is the well-known age--extinction degeneracy.}
when \lya\ is taken into account. 
Naively one may have expected the opposite, namely that the inclusion of \lya\ would 
lead to older age estimates, since for a fixed redshift, the contribution of \lya\ 
produces a bluer UV spectrum \citep[cf.][]{Finkelstein09}. This shows the importance 
of consistently fitting the physical parameters and redshift using the same spectral templates.

\subsubsection{Extinction}
Although a spread is obtained in $A_V$ for the different model assumptions,
the best-fit model values of $A_V$ correlate reasonably well around our ``reference value''
for most objects. The greatest differences arise when the SF history is varied
-- here imposing $\tau \ge 10$ Myr leads to lower $A_V$ and higher ages (because of the
age--extinction degeneracy), as shown by the yellow symbols -- and, for the same reason,
when nebular lines are omitted, as shown by some outliers with red symbols.
The largest differences in stellar mass are found for some objects where
nebular emission leads to a strong age reduction and hence a lower M/L ratio.

\subsubsection{SFR}
The star-formation rate SFR deserves special comment.
The value of the current SFR 
(SFR$(t)$) obtained from SED fits depends strongly on both SFH and age.
Formally, SFR=0 for instantaneous burst models, which are also considered here
and in other publications. Furthermore since the UV luminosity emitted
per unit SFR varies by $\sim 1$ order of magnitude within $\sim$ 100 Myr even
for constant star formation, the current SFR resulting from SED fits can 
differ\footnote{In general, one has SFR$(t) > $SFR(UV), since stars over a broad
age range contribute to the UV output, and since $L_{\rm UV}/$SFR is lower 
at young ages.}
from the usual SFR(UV) calibrations by 1 dex or more depending on both SF history and age.
For this reason, we compare in  Fig.\ \ref{fig_compare} the best-fit SFR values to 
SFR(UV), the value derived from the $J$ magnitude using the \citet{Kenn98} calibration,
and assuming no extinction.
The following differences can be seen: SFR $>$ SFR(UV) is obtained for many objects, where
$A_V>0$ and $t \ll$ 100 Myr.
Lower SFR$(t) <$SFR(UV) values are obtained for some objects with relative short
timescales $\tau$, i.e., rapidly declining SF histories.

The comparisons shown here all include a marginalisation over the three 
metallicities $Z$ considered in our models. As already mentioned, fixing $Z$
leads to small differences.
Given our limited knowledge of high-$z$ galaxies, other variations in the input physics, 
e.g., for the star-formation history or the reddening law, could be considered. 
Quantifying this is beyond the scope of this paper. The effect of 
different reddening laws on the physical parameters, however, can be understood 
quite simply, and is e.g., illustrated in \citet{Yabe09}.
Rising SF histories, which are not considered here, have, e.g., been advocated
by \citet{Finlator07}. We note, that these SF histories would in general 
correspond to a higher extinction than decreasing ones, to reproduce the observed 
SEDs of LBGs \citet[cf.][]{Finlator07}. This would strengthen our arguments
in favour of dust at $z \sim 7$.

\subsection{Correlations between physical properties in $z \sim 7$ galaxies}

Although the uncertainties in the physical parameters are relatively large for most 
objects, it is helpful to search for correlations between them, and
with observed quantities.
Figure \ref{fig_overview} shows the best-fit values for age, $A_V$, stellar mass, and 
SFR of all objects from the 3 samples, plotted as a function of the $J$-band magnitude,
which traces approximately the rest-frame UV at 1500 \AA.
The values plotted here are taken from our model 1, i.e., the reference 
model plus the constraint that $\tau \ge 10$ Myr, in particular to assure
that SFR is properly defined.
We also compare the best-fit values obtained by \citet{Gonzalez09} and \citet{Labbe10}
for the mean/stacked SED for the intermediate and faint samples, and the available
fit results from \citet{Capak09} for the bright objects. The origin of the 
differences with these authors have already been discussed above.

\begin{figure}[tb]
\includegraphics[width=8.8cm]{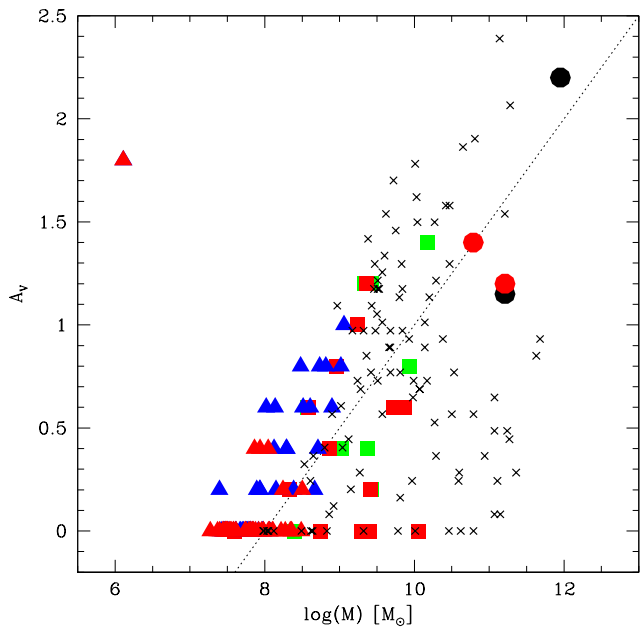}
\caption{Black circles, green squares, blue triangles show best-fit values of 
$A_V$ versus $M_\star$ derived from our reference model for ojects from the bright, 
intermediate, and faint samples respectively.
Red filled symbols show the same but from the reference
model imposing the restriction of $\tau \ge 10$ Myr (model 1).
The dotted line, given by $A_V = \log(M_\star /10^8)^{0.5}$, is shown to
guide the eye.
Small crosses show the fit results from \citet{Yabe09} at $z \sim 5$.
}
\label{fig_mass_av}
\end{figure}

As is clear for the different samples, stellar ages and 
extinction cover a wide range of values with no clear or strong trend. 
It is possible that a tendency of increasing $A_V$ for brighter objects exists, as 
would be expected from other studies at lower redshift (cf.\ below).
This trend becomes clearer when $A_V$ is plotted as a function of stellar mass,
as shown in Fig.\ \ref{fig_mass_av}.
We also find a tentative trend of the extinction $A_V$ with galaxy mass $M_\star$,
as shown in Fig.\ \ref{fig_mass_av}.

\begin{figure}[tb]
\includegraphics[width=8.8cm]{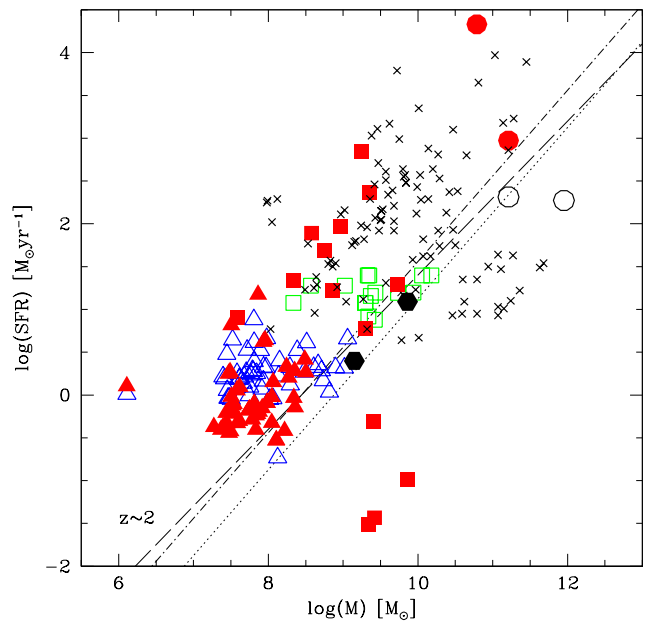}
\caption{Stellar mass--SFR relation for the $z \sim 7$ galaxies.
Open symbols (circles, squares, triangles) show the ``standard'' SFR(UV) value 
(not corrected for extinction) versus mass derived from our reference
model for objects from the bright, intermediate, and faint samples respectively.
Red filled symbols show the best-fit model SFR and $M_\star$ values 
when assuming $\tau \ge 10$ Myr (model 1).
Filled hexagons denote the mean values derived by \citet{Gonzalez09} and \citet{Labbe10}.
Small crosses show the fit results from \citet{Yabe09} at $z \sim 5$.
The dashed line shows the relation found by \citet{Daddi07} for $z \sim 2$ SF galaxies.
The dotted (dash-dotted) lines show the locus for SFR=const from $z=\infty$ (10) to 7 
corresponding to SSFR=1.3 (3.6) Gyr$^{-1}$. 
}
\label{fig_mass_sfr}
\end{figure}

Both the stellar mass and SFR show clear trends with the $J$-band magnitude, albeit
with a significant scatter. We note 
that the scatter in the mass-magnitude relation for the $z \sim 7$ objects 
does not significantly decrease when 3.6 or 4.5 \micron\ photometry is used.
Deviations of the best-fit SFR from the simple ``standard'' calibration, also indicated 
in Fig.\ \ref{fig_overview} for $A_V=0$,  are caused by non-zero extinction, age effects, 
and exponentially decreasing SF histories, or combinations thereof, as already discussed above.
Figure \ref{fig_mass_sfr} shows the corresponding correlation between stellar
mass and SFR, suggesting a fairly well-defined mass--SFR relation at $z \sim 7$.
Our best-fit values yield on average higher specific star-formation rates 
(SSFR$=$SFR$/M_\star$) than the values derived by \citet{Gonzalez09} and \citet{Labbe10}
shown by large open circles, and tend to indicate a relatively large
spread in SSFR.

\subsection{Comparison with lower redshift galaxies}
Although an exhaustive comparison with studies of the physical
properties of LBGs at lower redshift is clearly beyond the scope
of the present paper, it is worthwhile comparing briefly
our results with those obtained at $z \sim 5$, and at lower redshift.
To do so we refer to the detailed SED fitting analysis and discussions 
of \citet[][ and references therein]{Yabe09}, who have compared the physical parameters of
$z=2$, 3, 5, and 6 LBGs from different samples.

The analysis of both Yabe et al.\ and \citet{Verma07}
of $z \sim 5$ galaxies find 
clear signs of significant extinction. The former study finds $E(B-V)$ values ranging 
from 0 to $\sim$ 0.5, with a median of $E(B-V)-0.22$, corresponding to
$A_V=0.89$ i.e.\ a factor 9 attenuation of the UV flux!  
Our typical $A_V$ values are lower and the uncertainties are large;
but in many cases the best-fit model extinction is $A_V>0$ for the $z \sim 7$ LBGs.
For comparison, \citet{Bouwens09_beta,Bouwens09_betaz7} advocate
an attenuation of the UV flux by a factor 1.35--1.6 (corresponding
to $A_V \sim$ 0.12--0.19 for the Calzetti law) at $z \approx 7$, and
an attenuation by 2.7 ($A_V \sim 0.4$) at $z \approx 5$.
The extinction obtained by \citet{Yabe09} and from our analysis is thus typically
higher by a factor 3 than in the work of Bouwens and collaborators,
who estimate the dust attenuation from the UV slope.

We find a trend of increasing extinction with galaxy mass (Fig.\ \ref{fig_mass_av}).
At $z \sim$ 0--2, it is known that dust extinction increases with 
bolometric luminosity (i.e., also with SFR), which in turn increases with stellar 
mass \citep[e.g.,][]{Buat05,Buat08,Burgarella07,Daddi07,Reddy06,Reddy08}.
Luminosity-dependent dust corrections have been proposed, e.g., by
\citet{Reddy09} and \citet{Bouwens09_beta}, based on observed variations of the UV slope
with $M_{\rm UV}$. From these results, it is not surprising to find a
similar relation at $z \sim 7$, here expressed as extinction versus
stellar mass. The origin of a mass--extinction relation is most likely 
related to that of the mass--metallicity relation
\citep[cf.][]{Tremonti04,Erb06,Finlator07,Maiolino08}.
The data of \citet{Yabe09} at $z \sim 5$, also plotted in 
Fig.\ \ref{fig_mass_av}, may show a less clear trend with mass and some offset.
A systematic differential analysis of LBG samples at different redshifts,
will be necessary to shed more light on these issues.

The best-fit model values of the other physical parameters (age, mass, and SFR)
span a similar range as found by \citet{Yabe09} for $z \sim 5$ LBGs.
In contrast to the results of \citet{Gonzalez09} and \citet{Labbe10}, our relatively 
young ages resemble those found by \citet{Verma07} and \citet{Yabe09} for $z \sim 5$, 
which are younger than $z \sim$ 2--3 LBGs \citep[cf.][]{Shap01,Shap05,Sawicki07,
Yabe09}. However, other studies \citep[e.g.,][]{Stark07_massdensity,Eyles07,Stark09}, find relatively
old ages at $z \sim 6$, and \citet{Yabe09} confirm some of them with 
their method. Whether LBGs show a clear trend of decreasing age with 
increasing redshift, as one may naively expect, thus remains to be clarified.

For the IRAC-detected objects at $z\sim 7$ (the bright and intermediate 
samples), the relation between stellar mass 
and absolute optical magnitude (derived from the 4.5 \micron\ flux) 
is very similar to that of the $z \sim 5$ objects. The lower masses of the 
faint $z \sim 7$ subsample -- undetected by IRAC -- provide a natural
extension towards fainter objects. 
A similar behaviour is also found for the SFR, when comparing our
$z \sim 7$ results to those of \citet{Yabe09}.

The $M_\star$--SFR relation found at $z \sim 7$ resembles that of  $z \approx 5$ LBGs
(see Fig.\ \ref{fig_mass_sfr}).
We note that most objects lie above the $M_\star$--SFR relation found
at $z\approx 2$ \citep[cf.][]{Daddi07,Sawicki07}. 
However, both the behaviour and the scatter found for the $z \sim 7$ objects resembles 
that of $z \approx 5$ LBGs, also shown in this plot.
Given the large uncertainties for the $z\sim 7$ objects, and the different
methods used in these analysis, it is possible that the same  $M_\star$--SFR
relation (with a similar scatter) is in place from $z \sim 7$ to 2
and that the specific SFR declines to lower redshift
\citep[cf.][]{Daddi07,Elbaz07,Noeske07,Gonzalez09}.

A detailed study of the physical parameters of LBGs over a wide
redshift domain with our modeling tools, including more in-depth comparisons
of the results from different methods/groups, will be presented elsewhere.

\subsection{Implications}
An important result of this study is the possible discovery of dust extinction
in LBGs at $z \sim 7$ from detailed SED modeling, which contrasts with claims of 
basically dust-free objects from studies of their UV slope or from too
restrictive SED modeling \citep[cf.][]{Bouwens09_betaz7,Gonzalez09,Labbe10}.
If this were correct, the UV attenuation -- up to $A_V \sim 1.2$ (factor 20) 
for the brightest objects -- would imply a higher SFR 
density at high redshift than previously inferred.
However, to quantify the average dust correction remains difficult, especially 
given the large uncertainties for the individual objects and the possible
dependence of $A_V$ on the UV luminosity and/or stellar mass.
Even if our results imply that a non-negligible dust correction
is required at $z \sim 7$, it is possible that the SFR density decreases with
redshift from $z \ga$ 3--5 on, at least if the median attenuation of 
$A_V \sim 0.9$ found by \citet{Yabe09} at $z \approx 5$ is representative.

Our analysis, covering a wider domain in parameters space than previous studies
and also allowing for the effects of nebular emission, shows that a wide range 
of stellar ages is acceptable for most $z \sim 7$ LBGs.
The Balmer break observed for some objects does not always imply old ages,
or correspondingly high formation redshifts as already pointed out 
by \citet{SdB09}. It can often equally well
be explained by nebular emission, younger populations at non-solar
metallicity, extinction, or a combination thereof.

Younger ages and variations in the star-formation history can lead to solutions
with lower masses and higher SFR for many of the $z \sim 7$ LBGs than
estimated by other groups \citep{Gonzalez09,Labbe10}.
The specific star-formation rate (SSFR$=$SFR$/M_\star$) of
these high-$z$ galaxies may also be higher than thought.

In short, our SED analysis including in particular the effects
of nebular emission and considering a wide parameter space 
(SF histories, extinction etc., shows that the physical 
parameters of $z \sim 7$ galaxies may differ significantly 
from those advocated by other groups, with possibly many 
implications for our picture of galaxy formation and evolution
in the early universe. 
However, we emphasize that the physical properties of 
these objects are affected by large uncertainties (cf.\ Sect.\
\ref{s_res} and Appendix), given their faintness and the available data.
Furthermore, comparisons with results for lower redshifts may 
be problematic because of the use of different methods.

Obviously, additional and higher quality data, and a detailed differential study 
of the physical parameters of LBGs over a wide redshift domain using state-of-the-art
tools will be very helpful in providing a clearer and more accurate picture
of the properties and evolution of high-$z$ galaxies in the distant universe, 
and their link to lower redshift galaxy populations.

\section{Conclusions}
\label{s_conc}
We have presented a homogeneous and detailed analysis of broad-band
photometry for three samples of $z \sim$6--8 galaxies discovered
by the COSMOS survey and with HST 
\citep[see][]{Capak09,Gonzalez09,Oesch09,Mclure09, Bunker09}.  Their $J$-band
magnitude span a range from $\sim$ 23 to 29, the bulk of them having
$J_{AB} \sim$ 26--29.  The broad-band SEDs have been fitted using our
modified version of the \hyperz\ photometric redshift code
described in \citet{SdB09}, which accounts for the effects of nebular
emission.

In contrast to earlier studies that assumed, e.g., constant 
star formation and/or no extinction, we have explored a wide
range of the parameter space without using priors. 
The free parameters (and range) of our SED fits are:
the redshift $z$ ($[0,12]$),
the metallicity $Z$ ($Z/\zsun=$1, 1/5, 1/20),
the SF history described by the e-folding timescale $\tau$  
($[0,1000]$ Myr, $\infty$),
the age $t$ defined since the onset of star-formation ($\le t_H$), 
the extinction $A_V$ ($[0,2]$ in general) described by the Calzetti law \citep{Calzetti00},
and whether or not nebular emission is included
\footnote{In some cases, we exclude the \lya\ line from the synthetic 
spectra, since this line may be attenuated by radiation transfer processes
inside the galaxy or by the intervening intergalactic medium.}.

Our main results can be summarised as follows:
\begin{itemize}
\item Overall, we find that the physical parameters of most 
galaxies studied here cover a wide range of acceptable values, e.g., within
$\Delta \ki2 \la 1$ from the best-fit modeling. This finding is independent of 
whether nebular emission is included or not.

\item Stellar ages, in particular, are not tightly constrained, even for objects
detected with Spitzer, i.e., with photometry both blue- and redward of 
the Balmer break. When nebular emission is taken into account, we find
that the majority of the objects (and the stacked SEDs as well) are 
most accurately reproduced by
ages $t < 100$, which are younger than derived in other studies of the same objects
\citep{Gonzalez09,Labbe10}. The younger ages are due to the contribution of 
nebular lines to the broad-band rest-frame optical filters, which mimic
to some extent a Balmer break, as already shown by \citet{SdB09}. 

\item Examination of the UV slope and SEDs of faint z-dropout
galaxies found with WFC3/HST shows no need for ``unusual''
stellar populations, extreme metallicities, or other physical processes,
advocated previously by \citet{Bouwens09_betaz7}, when the uncertainties are 
taken into account \citep[see also][]{Finkelstein09}.

\item Albeit with large uncertainties, our fit results show indications
of dust attenuation in some of the $z \approx$ 6--8 galaxies, with
best-fit model values of $A_V$ up to $\sim$ 1, even among relatively faint
objects ($J_{AB} \sim$ 26--27; cf.\ Fig.\ \ref{fig_overview}).

\item We find a possible trend of increasing dust attenuation with 
the stellar mass of the galaxy (Fig.\ \ref{fig_mass_av}) and a 
relatively large scatter in specific star-formation rates, SFR/$M_\star$.

\item Our results, including the evidence of dust in $z \approx$ 6--8
galaxies, are consistent with the results and trends from other SED studies 
of LBG samples at $z \sim 5$ \citep[see e.g.][]{Verma07,Yabe09}.
\end{itemize}

We will present elsewhere a systematic study of the evolution of the
physical parameters of LBGs at different redshifts, adopting a
homogeneous method and including a detailed error analysis.

\acknowledgements
We would like to thank Matthew Hayes, David Valls-Gabaud, 
Rychard Bouwens, and Masami Ouchi for interesting discussions,
and Roser Pell\'o for regular discussions and support with \hyperz.
This work was supported by the Swiss National Science Foundation.
\newpage
\appendix

\section{Bright sample \citep{Capak09}}
\begin{figure}[tb]
\includegraphics[width=8.8cm]{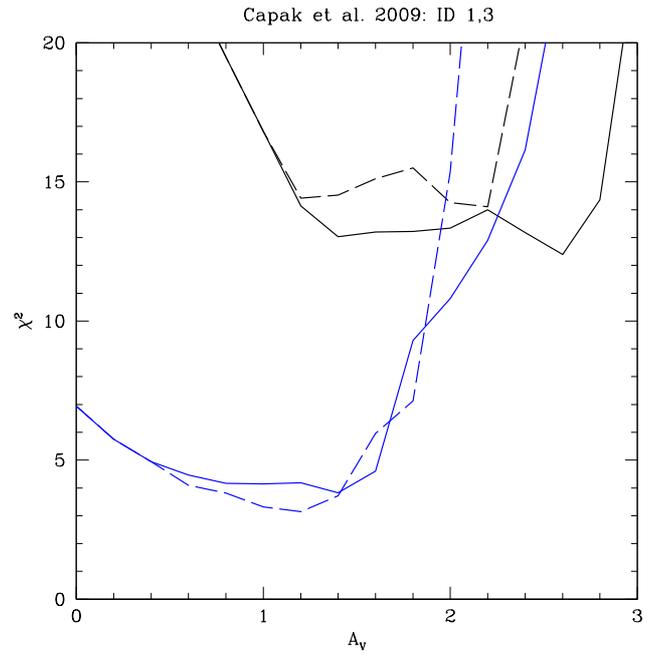}
\caption{Distribution of the minimum \ki2\ value as a function of the 
attenuation $A_V$ for object 1 (black lines) and 3 (blue lines) from
the bright sample \citep{Capak09}.
Solid lines show \ki2\ obtained with standard spectral templates (neglecting
nebular emission), dashed lines for templates including nebular emission,
but with \lya\ suppressed.}
\label{fig_capak_av}
\end{figure}
\begin{figure}[tb]
\includegraphics[width=8.8cm]{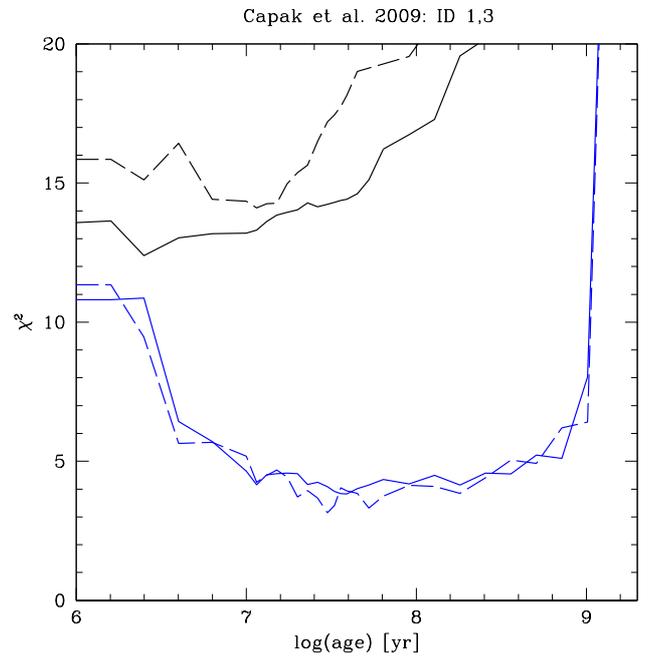}
\caption{Same as Fig.\ \protect\ref{fig_capak_av} as a function of stellar age.}
\label{fig_capak_age}
\end{figure}
\begin{figure}[tb]
\includegraphics[width=8.8cm]{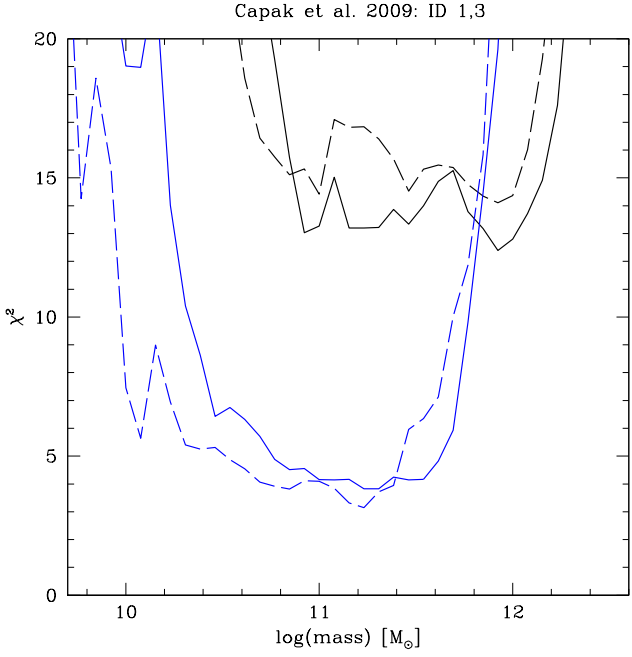}
\caption{Same as Fig.\ \protect\ref{fig_capak_av} as a function of stellar mass.}
\label{fig_capak_mass}
\end{figure}
\begin{figure}[tb]
\includegraphics[width=8.8cm]{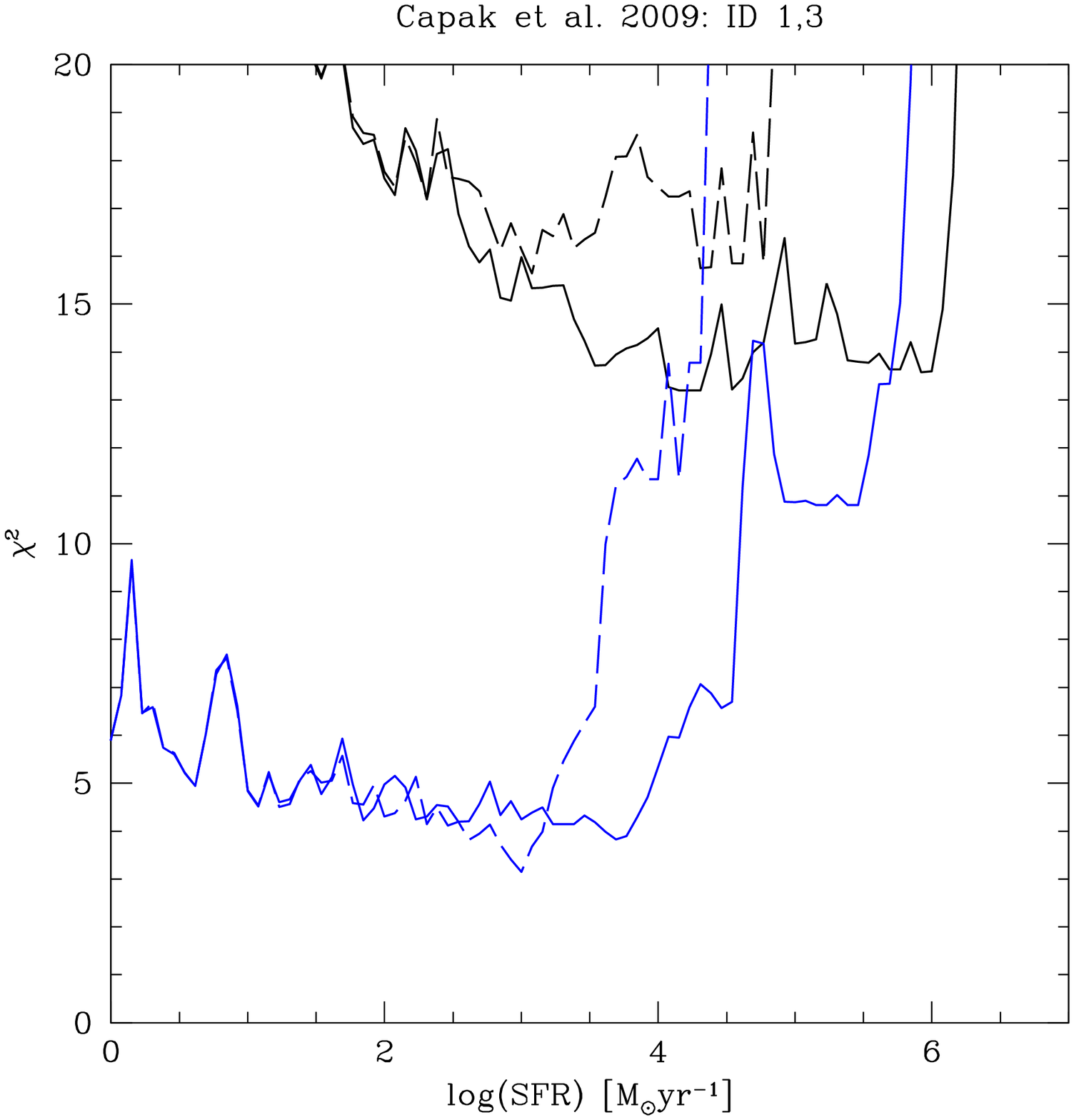}
\caption{Same as Fig.\ \protect\ref{fig_capak_av} as a function of the star-formation rate.}
\label{fig_capak_sfr}
\end{figure}

\section{Intermediate sample \citep{Gonzalez09}}
\begin{figure}[tb]
\includegraphics[width=8.8cm]{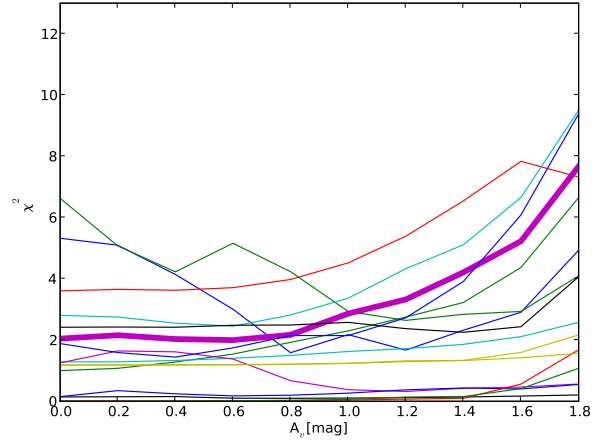}
\caption{Distribution of the minimum \ki2\ value as a function of the 
attenuation $A_V$ for all objects from the ``intermediate sample''
of \citet{Gonzalez09} computed with the spectral templates including 
nebular emission, but with \lya\ suppressed.. 
Different colours represent different objects.
The thick purple line show \ki2\ for the mean SED.}
\label{fig_gonz_chi2_av}
\end{figure}

\begin{figure}[tb]
\includegraphics[width=8.8cm]{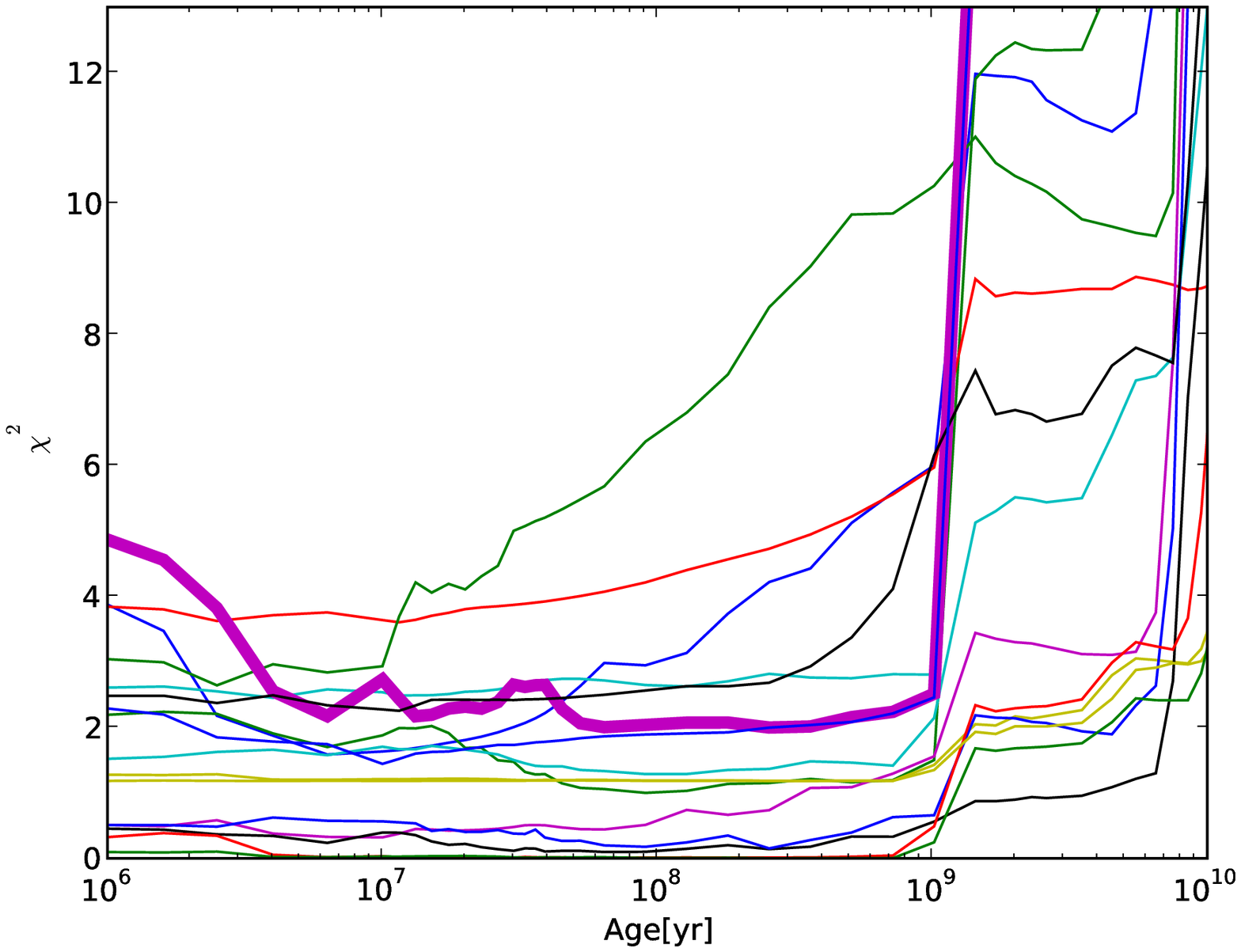}
\caption{Same as Fig.\ \protect\ref{fig_gonz_chi2_av} as a function of the stellar age.}
\label{fig_gonz_chi2_age}
\end{figure}

\begin{figure}[tb]
\includegraphics[width=8.8cm]{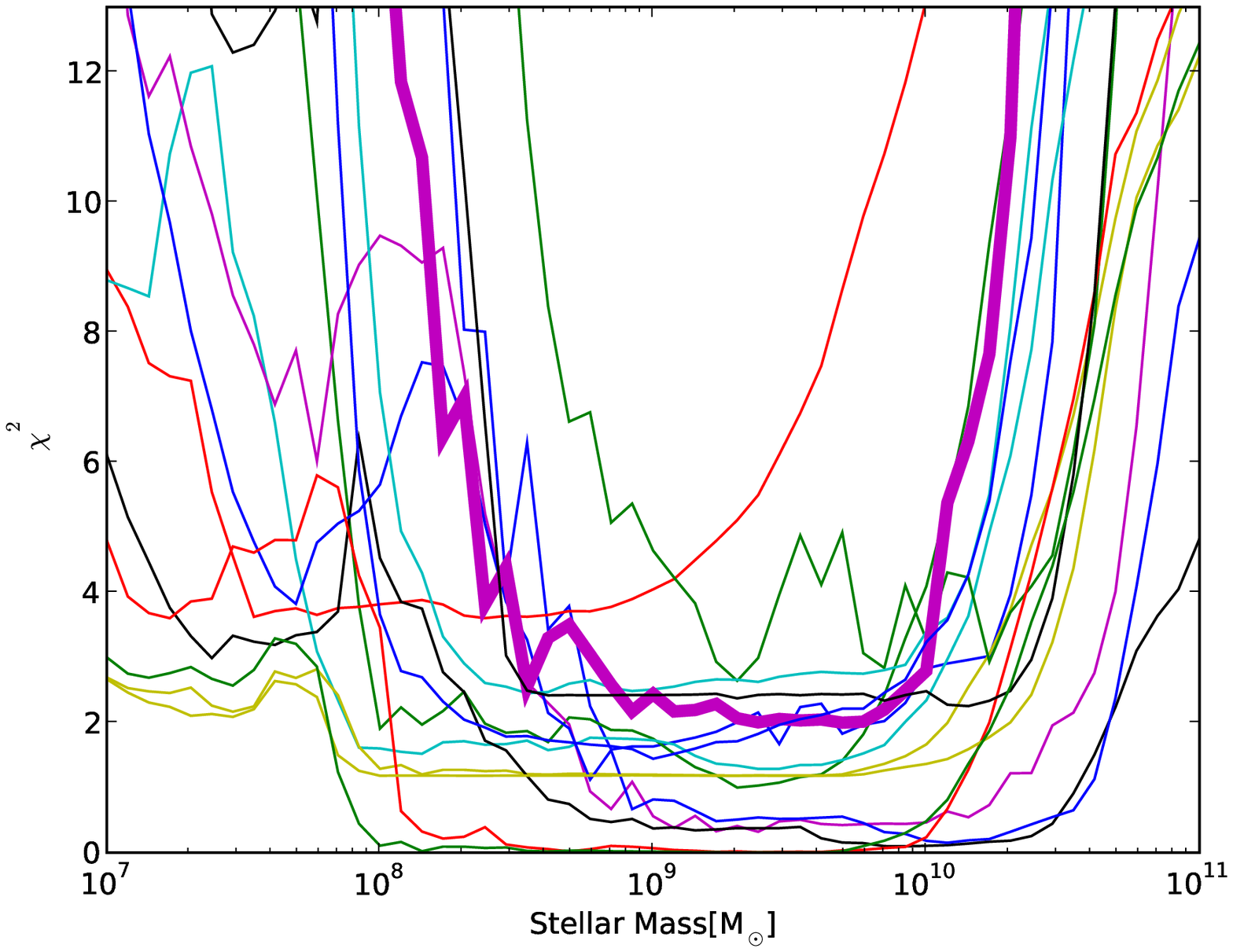}
\caption{Same as Fig.\ \protect\ref{fig_gonz_chi2_av} as a function of the stellar mass.}
\label{fig_gonz_chi2_mass}
\end{figure}

\begin{figure}[tb]
\includegraphics[width=8.8cm]{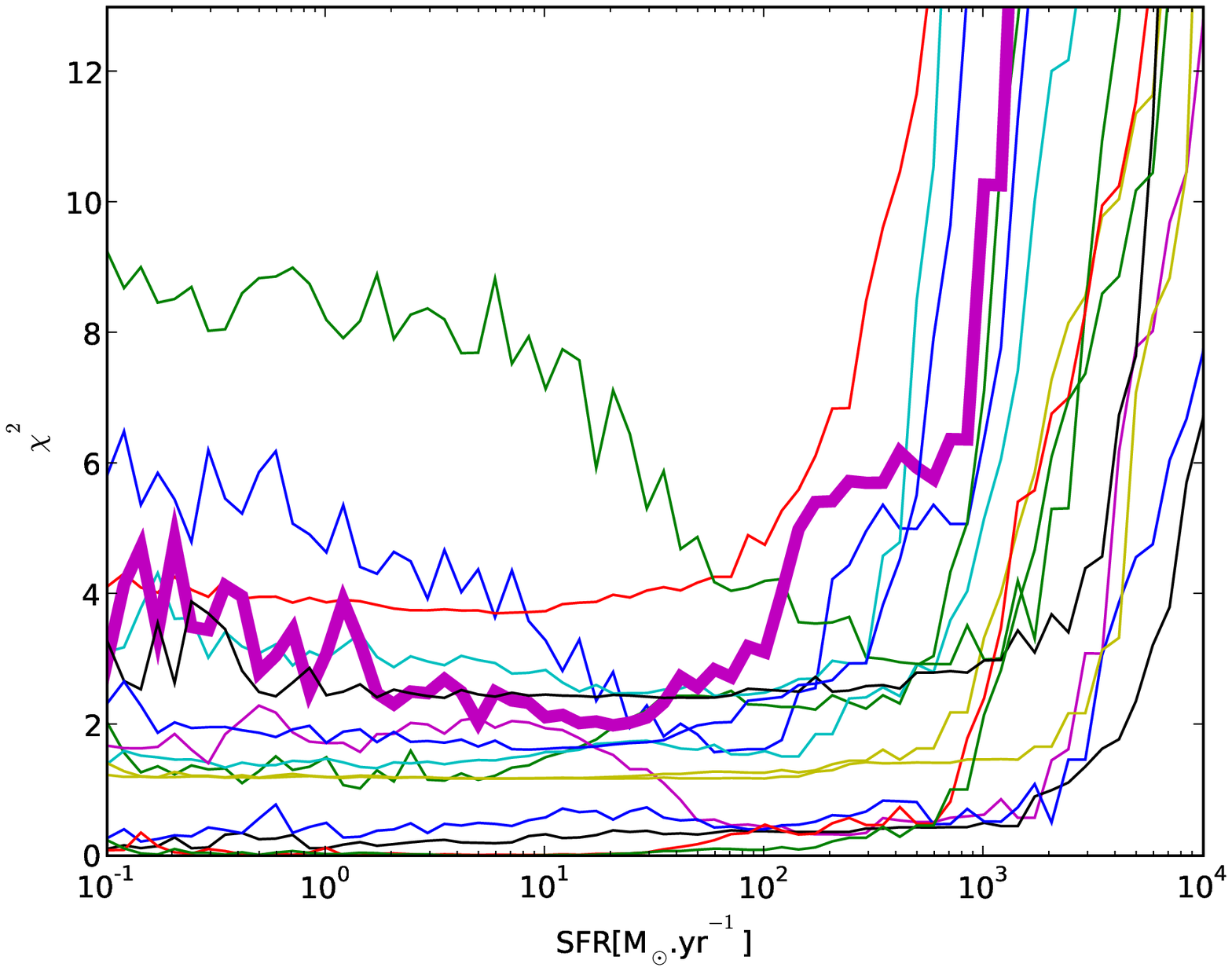}
\caption{Same as Fig.\ \protect\ref{fig_gonz_chi2_av} as a function of the star-formation rate.}
\label{fig_gonz_chi2_sfr}
\end{figure}

\section{Faint sample}
\subsection{\citet{Oesch09} objects}

\begin{figure}[tb]
\includegraphics[width=8.8cm]{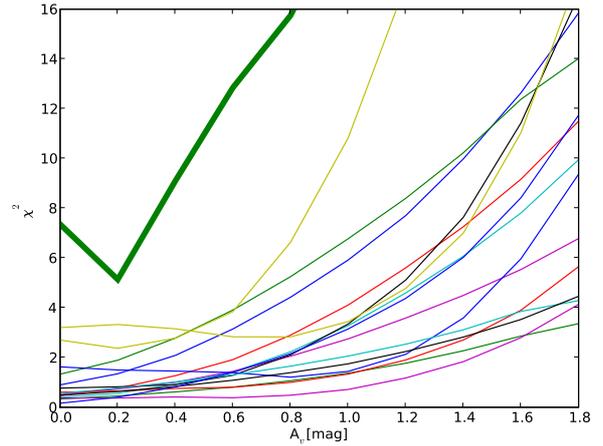}
\caption{Distribution of the minimum \ki2\ value as a function of the 
attenuation $A_V$ for the 15 objects from the ``faint sample''
of \citet{Oesch09} computed with the spectral templates including 
nebular emission, but with \lya\ suppressed.. 
Different colours represent different objects;
the thick green line for the stack of 14 objects of the same sample
measured by \citet{Labbe10}.}
\label{fig_mclure_chi2_av}
\end{figure}

\begin{figure}[tb]
\includegraphics[width=8.8cm]{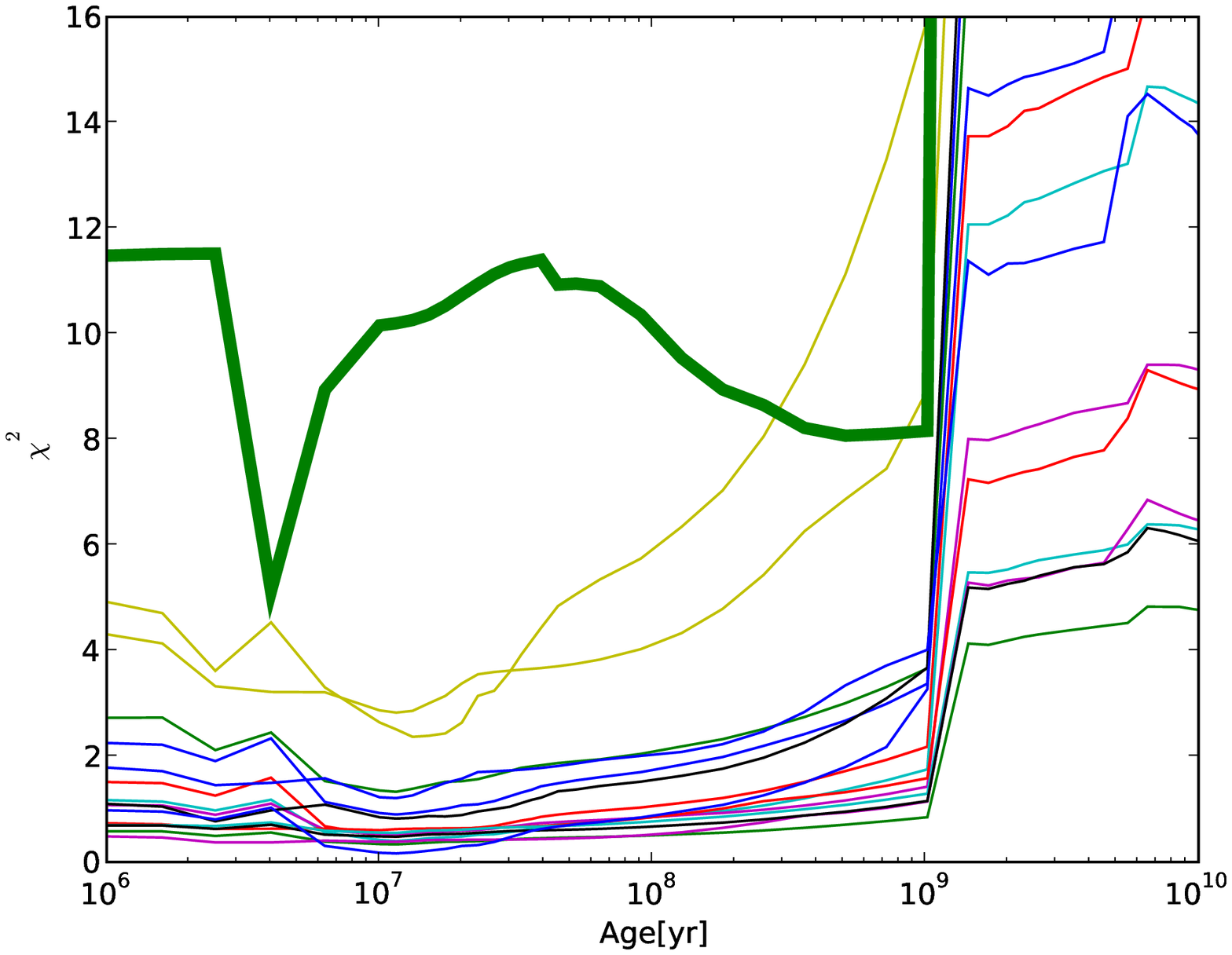}
\caption{Same as Fig.\ \protect\ref{fig_mclure_chi2_av} as a function of the stellar age.}
\label{fig_mclure_chi2_age}
\end{figure}

\begin{figure}[tb]
\includegraphics[width=8.8cm]{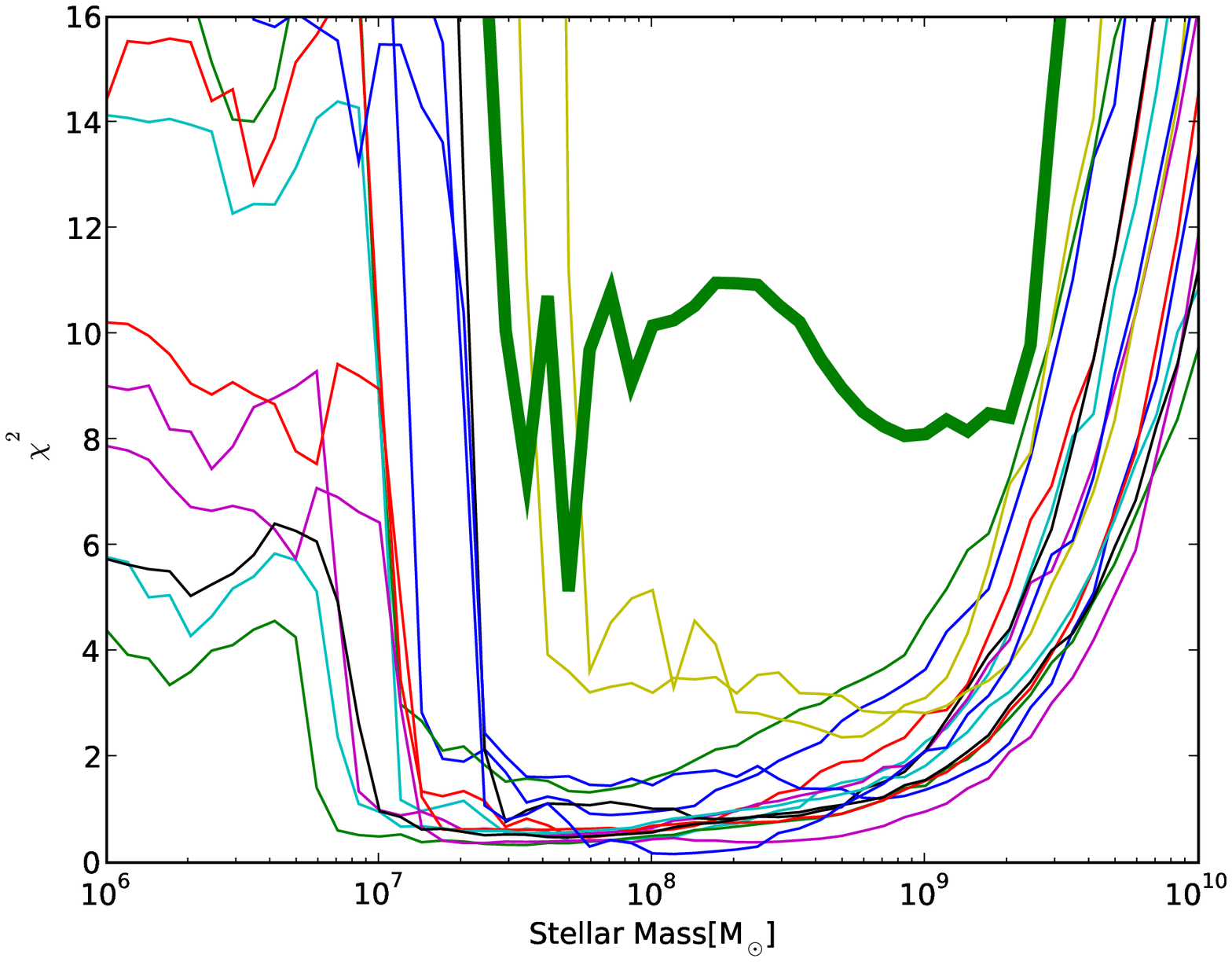}
\caption{Same as Fig.\ \protect\ref{fig_mclure_chi2_av} as a function of the stellar mass.}
\label{fig_mclure_chi2_mass}
\end{figure}

\begin{figure}[tb]
\includegraphics[width=8.8cm]{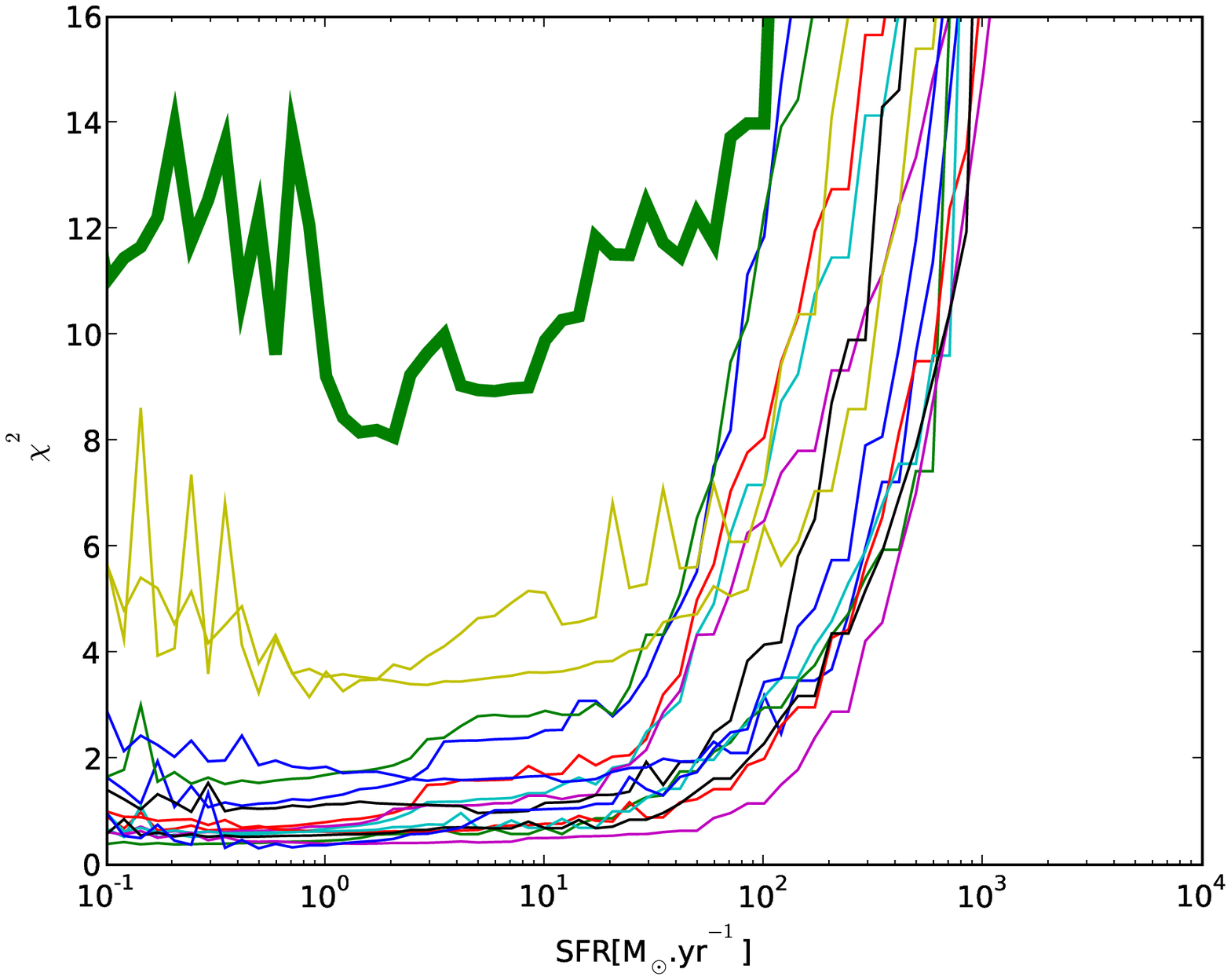}
\caption{Same as Fig.\ \protect\ref{fig_mclure_chi2_av} as a function of the star-formation rate.}
\label{fig_mclure_chi2_sfr}
\end{figure}

\subsection{Stacked SED \citep{Labbe10}}
\begin{figure}[tb]
\includegraphics[width=8.8cm]{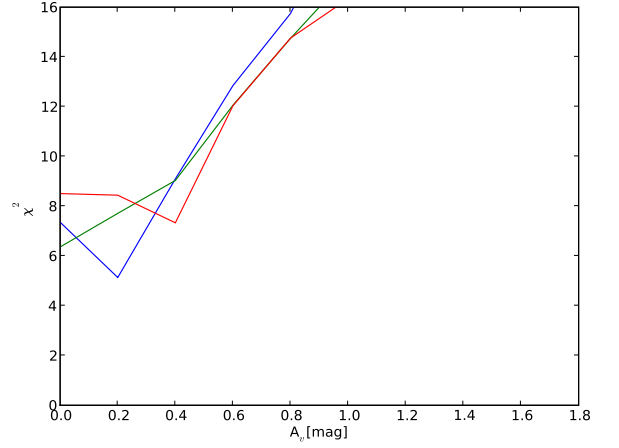}
\caption{Distribution of the minimum \ki2\ value as a function of the 
attenuation $A_V$ for the stack of 14 objects from the ``faint sample''
of \citep{Labbe10}.
Different colours show the results from
fits using spectral templates including nebular emission (green), 
nebular emission but \lya\ suppressed (blue), and 
no nebular emission (red).}
\label{fig_stack_av}
\end{figure}

\begin{figure}[tb]
\includegraphics[width=8.8cm]{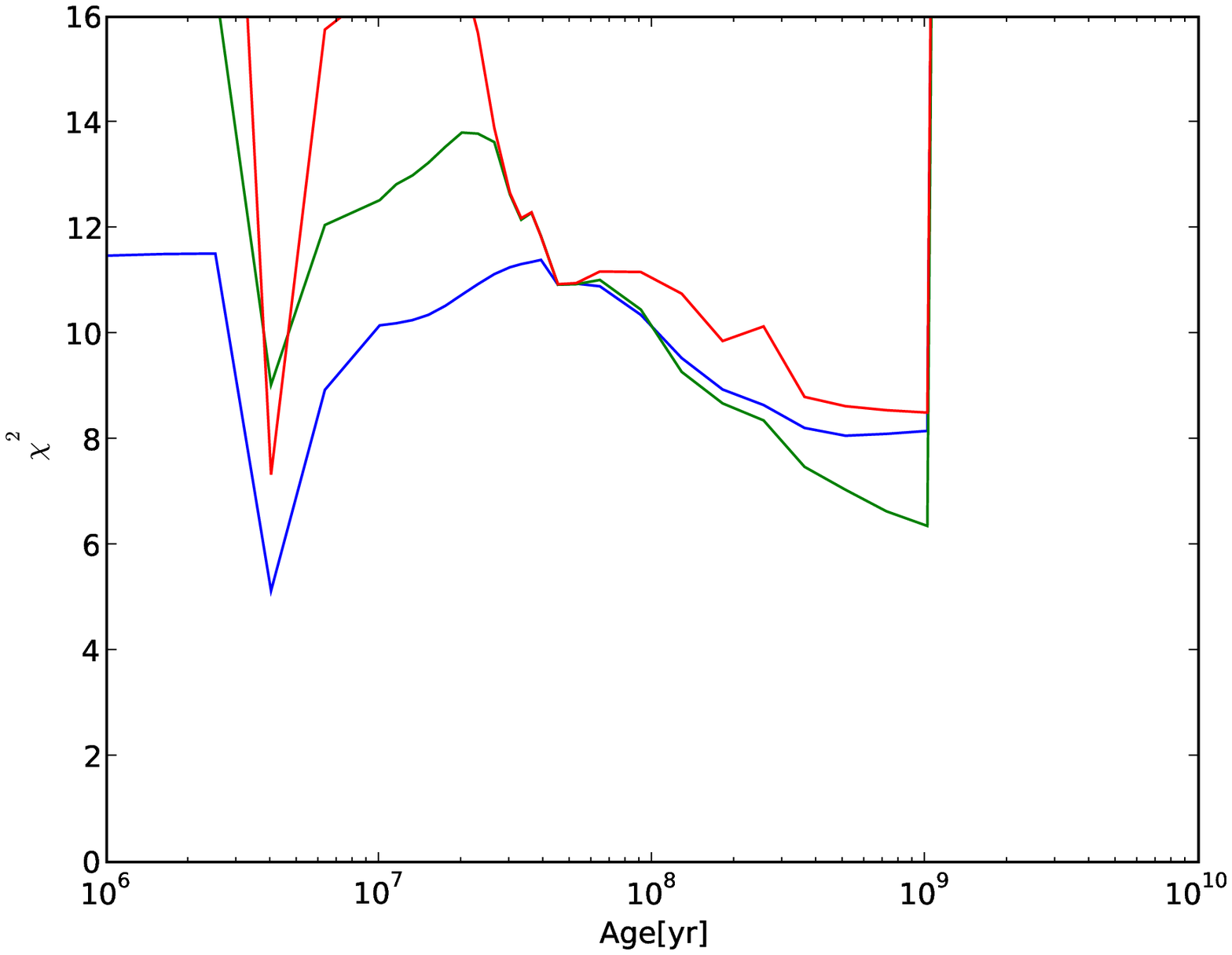}
\caption{Same as Fig.\ \protect\ref{fig_stack_av} as a function of the stellar age.}
\label{fig_stack_age}
\end{figure}

\begin{figure}[tb]
\includegraphics[width=8.8cm]{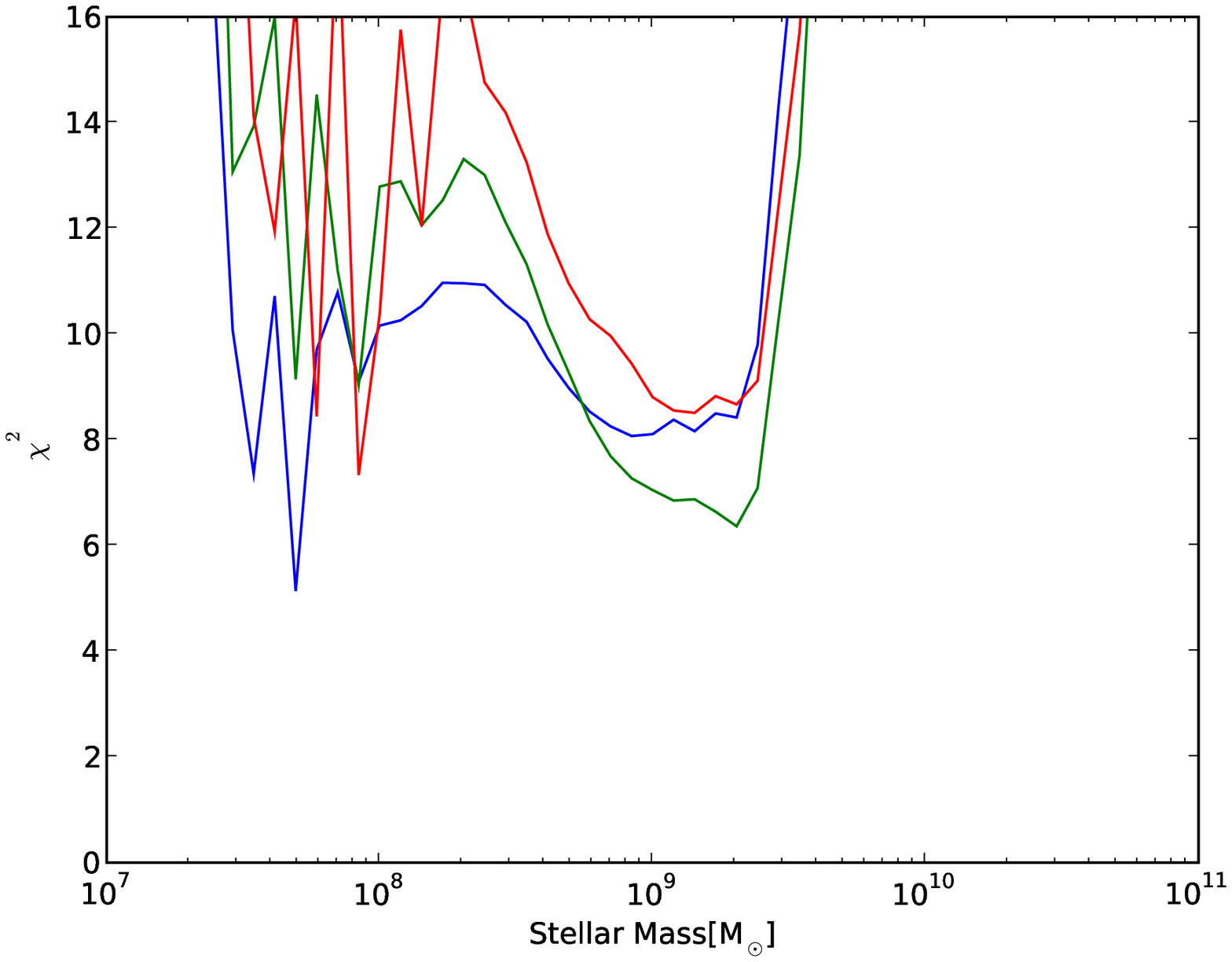}
\caption{Same as Fig.\ \protect\ref{fig_stack_av} as a function of the stellar mass.}
\label{fig_stack_mass}
\end{figure}

\begin{figure}[tb]
\includegraphics[width=8.8cm]{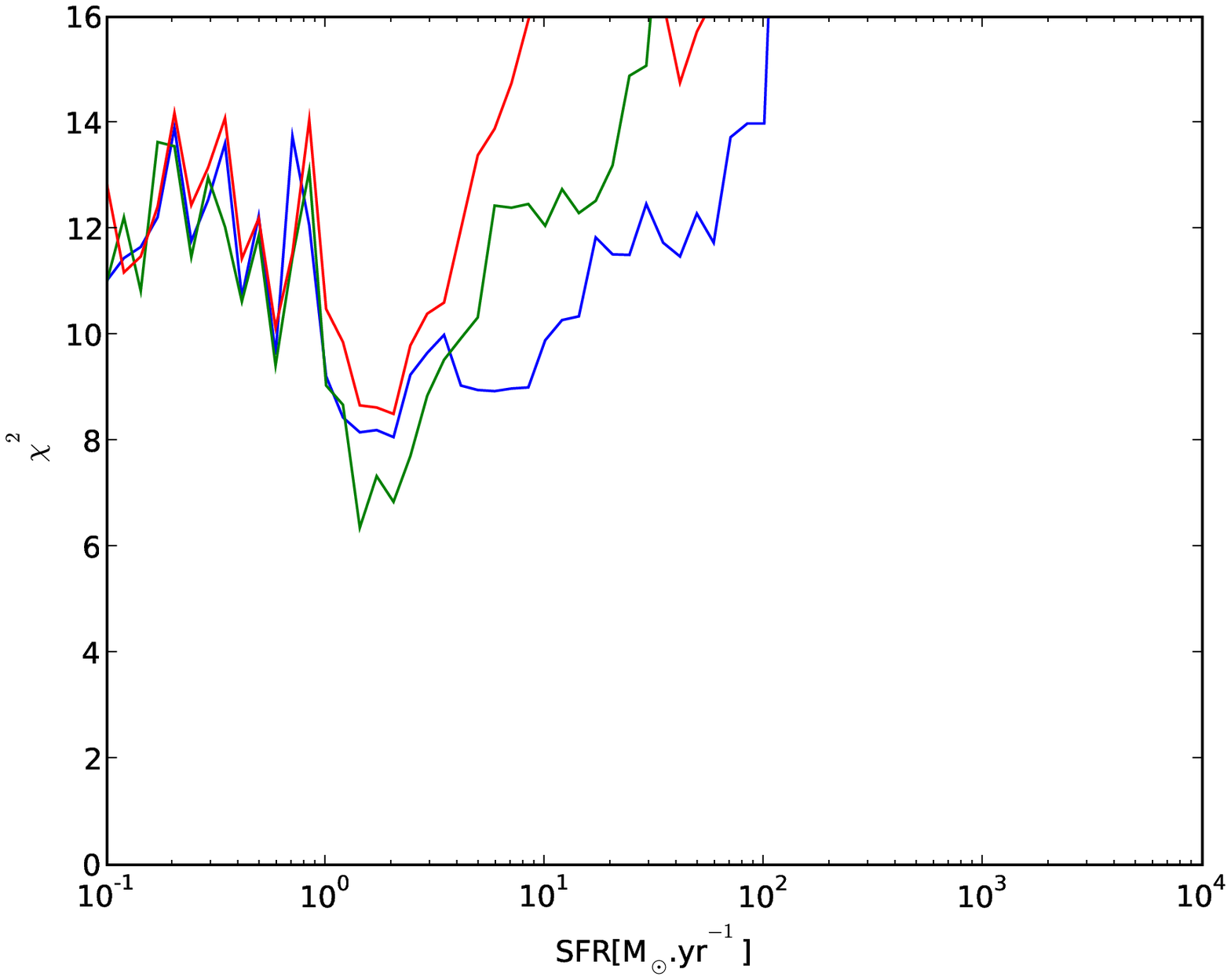}
\caption{Same as Fig.\ \protect\ref{fig_stack_av} as a function of the star-formation rate.}
\label{fig_stack_sfr}
\end{figure}

\bibliographystyle{aa}
\bibliography{references}

\end{document}